\documentclass[nologo,11pt,a4paper,hidelinks]{ETHpaper}

\usepackage{graphicx}
\usepackage[square,numbers,sort&compress]{natbib}

\usepackage{amsmath}
\usepackage{amssymb}
\usepackage{amsthm}

\makeatletter
\let\cl@part\@undefined
\makeatother

\usepackage[capitalise]{cleveref}
\crefname{thm}{Theorem}{Theorems}
\newcommand{\corref}[1]{Corollary~\ref{#1}}
\usepackage{dsfont}
\usepackage{url}

\newcommand{\abs}[1]{\left\lvert#1\right\rvert}
\DeclareMathOperator*{\argmax}{arg\,max}

\graphicspath{{./figures/}}

\DeclareMathOperator{\sgn}{sgn}
\newtheorem{defn}{Definition}
\newtheorem{thm}{Theorem}
\newtheorem{cor}{Corollary}

\title{The role of administrative costs in policies to improve cooperation}
\titlealternative{The role of administrative costs in policies to improve cooperation}

	\author{Giacomo Vaccario, Luca Verginer, Frank Schweitzer}
	\authoralternative{Giacomo Vaccario, Luca Verginer, Frank Schweitzer}%

	\address{ETH Zurich, Switzerland\\ \url{www.sg.ethz.ch}}

\www{\url{http://www.sg.ethz.ch}}

\reference{(preprint, to appear in \emph{Automatica})
}
\makeframing

\begin{document}
\maketitle

	\begin{abstract}
		We use an agent-based model for studying policies of an abstract government to increase the level of cooperation.
		In situations where defection is the rational choice, agents receive a bonus if they choose to cooperate.
		The government taxes agents' payoffs from a repeated prisoner's dilemma game to establish a budget that covers the payment of bonuses, but also the administrative costs involved.
		We consider costs to identify cooperators, and costs to mitigate the consequences of defective behavior.
		Governmental interventions are fallible, i.e.
		cooperators and defectors are correctly identified only with a certain accuracy.
		Hence, the critical bonus needed to let agents cooperate depends on the accuracy, but also on the current level of cooperation.
		We analyze the cost constraints for the government to pay the critical bonus and different policies to adjust the accuracy dependent on the level of cooperation.
		We find that perfect cooperation becomes impossible. Instead, high levels of cooperation can be achieved with optimal adjustment policies.
	\end{abstract}

\section{Introduction}

At a time where individuals' behavior can be effectively monitored by means of advanced technologies, it becomes tempting for governments or other central authorities, to utilize this information for steering  preferred actions, notably cooperation.
Experiments have shown that interventions can indeed lead to an increase of cooperation
\cite{Rand-Dreber-ea-2009-positiveinteractions,fehr2000cooperation,Ledyard-2020-publicgoods,Balliet-Mulder-Van-2011-reward}.
But this hinges on the accurate identification of cooperators and defectors, which is often fallible.
Punishing or rewarding the wrong targets, however, will have negative consequences.
Individuals can loose trust in the government's measures to improve cooperation.
They will also less likely choose cooperation once they learned that their expected reward does not appear.
Because identifying and incentivizing cooperative behavior is costly, governmental measures will find acceptance only if they can demonstrate their effectiveness in improving cooperation.

In this paper we will not discuss the practical or ethical issues of governmental interventions \citep{Yuan-Chen-2025-holdingai,Buchanan-2008-autonomypaternalismjustice}.
Instead we focus on a theoretical approach to address a research question that is often overlooked, the role of costs in intervention policies.
Specifically, we propose a model to formalize the subtle balance between incentives for cooperation, the costs associated with them, and the fallibility of measures implemented.
Our model highlights under which conditions cooperation can be efficiently induced, by analyzing the role of identification costs and the probability of correct identification.

From a broader theoretical perspective, we contribute to the long-lasting discussion how to improve cooperation in situations where defection is the rational choice of utility maximizing individuals.
Agent-based models have shown that group interactions \citep{Pena-2011-groupsize} and spatial proximity \citep{Weidenholzer-2010-coordinationgames, Schweitzer-Behera-Muhlenbein-2002-evolutioncooperation} can lead to higher levels of cooperation.
Also social herding \citep{Schweitzer-Mavrodiev-Tessone-2013-how,Vicario-2021-imitationlocalinteractions}, optimal migration \citep{Schweitzer-Behera-2012-optim,Helbing-Yu-2008-migrationas,Funk-Hauert-2019-directed,Jiang-Wang-ea-2010-role}, selection biases and partner matching \citep{Nax-Murphy-Helbing-2017-nashdynamics} support cooperation.

Other agent-based models study institutional incentives  \citep{Flores-Han-2024-evolution,Dong-Zhang-Tao-2016}, punishments or rewards
\citep{Ho-Luh-Olsder-1982, kanazawa2009replicator,Yang-Zhang-ea-2018-endogenous,Schweitzer-Verginer-Vaccario-2020-should}.
Instead of punishment, which has a negative connotation we focus on the positive side, namely to reward cooperating individuals by paying a \emph{bonus}.
But different from other approaches~\cite{cuesta2008,jimenez2009shared}, we study the problem that these bonuses are costly and therefore need to be financed in a self-sustained manner.

In our model, this task is assigned to the ``government'', a term we use as an abstraction of a central authority.
It has to secure funds to reward cooperators, but to also cover administrative costs associated with it.
Similar to \cite{Lorenz-Paetzel-Schweitzer-2013-redistribution,Schweitzer-Verginer-Vaccario-2020-should}, we introduce a taxation mechanism to establish a budget for the government.
This budget is spend on paying bonuses, but also on technologies to identify cooperators most accurately.
Further, we consider \emph{mitigation} costs  to compensate for the absence of cooperation.
This implies that there is a added value to cooperation.
Defection is not without consequence; this behavior causes damage to the system that the government must address.
Prominent examples are environmental protection, public health or sustainable resource management.
If nobody cooperates on these issues, governments have to invest their budgets to mitigate the consequences from defecting behavior.

That means, in our model the government plays the most important role.
It has to ensure a sufficient budget for incentivizing cooperation, but it also needs methods for identifying cooperators, on the one hand, and policy arguments to convince the public, on the other hand.
This in the end requires the government to establish the most effective policies to reach the goal.
A formal model to test some of these policy decisions therefore is a timely contribution.

The prominent role of the government builds upon and extends two important strands of research. 
The first involves game-theoretic models with environmental feedback \citep{gong2022limit, weitz2016oscillating, tilman2020evolutionary}, where cooperation levels dynamically influence and are influenced by a shared environment. 
In our setting, the government budget used to fund bonuses for cooperative behavior plays a role analogous to the environment. 
However, in contrast to passive environmental feedback, our model features an active government that not only responds to cooperation levels but also implements targeted intervention policies to stabilize desirable system dynamics. 
The second strand concerns subsidy-based control in multi-agent systems, typically modeled via replicator dynamics\citep{morimoto2015subsidy, li2022replicator, morimoto2012multipopulation, kanazawa2009replicator}. 
Like these studies, we aim to stabilize cooperation through subsidies, but we generalize the framework by moving beyond replicator equations. 
In our model, agents are boundedly rational, and replicator dynamics arise only as a special case. 
Indeed, we show that achieving cooperation is significantly easier under replicator dynamics than under more realistic assumptions of heterogeneous agent decision-making, which further highlights the necessity of our broader modeling approach.

Our paper is organized as follows:
In Section 2 we introduce an agent-based model that allows to formally address the two different perspectives on the problem, the one of utility-maximizing agents and the one of the government maximizing cooperation.
Specifically, we utilize the game-theoretical setting of a prisoner's dilemma, combined with a proportional taxation of the payoffs.
Budget considerations that include costs for identification and for mitigation of defectors then lead us to derive minimal and maximal bonus payments.
In Section 3 we study possible equilibria of the cooperation dynamics and compare the role of rational and bounded rational decisions on the level of cooperation.
Sections 4 and 5 analyze the impact of different costs in reaching a stable level of cooperation.
Using the no-cost scenario as our reference case, we first investigate mitigation costs, then identification costs and their combination.
In particular, we discuss different policy decisions to adjust the identification accuracy and their role in reaching a higher, and possibly the maximum affordable level of cooperation.

\section{Model}
\label{sec:model}

\subsection{The agent perspective}
\label{sec:agents}

\paragraph*{Expected cooperation.}
Consider a population of agents $i=0,...,N$ that interact in a game-theoretical setup, precisely in a repeated Prisoner's Dilemma (PD) game.
That means, agents' strategies $s_{t}^{i}\in\{0,1\}$ can be either to cooperate, $s^{i}_{t}=1$, or to defect, $s^{i}_{t}=0$.
Because agents can change their strategy over time, we define $p[s^{i}_{t}]$ as the probability of agent $i$ to use strategy $s^{i}_{t}$ at a discrete time step $t$, where $\Delta t=1$.
The \emph{expected fraction} of cooperators is then defined as:
\begin{align}
	\label{eq:5}
	\mathds{E}[f_{t}]\equiv f_{t} =\frac{1}{N}\sum_{i=1}^{N}p[s^{i}_{t}=1] =\frac{\mathds{E}[N^{c}_{t}]}{N}
\end{align}
where $\mathds{E}[N^{c}_{t}]\equiv N^{c}_{t}$ is the expected total number of cooperators at time step $t$ and $\mathds{E}[N^{d}_{t}]\equiv N^{d}_{t}=N-N^{c}_{t}$ the expected number of defectors.
To ease the notation, in the following we will use $f_{t}$, $N^{c}_{t}$ and $N^{d}_{t}$ for the \emph{expected} values.

\paragraph*{Payoffs in a PD game.}
In general interactions in the agent population can also include \emph{group interactions} \citep{Papanikolaou-Vaccario-ea-2022-consensus, papanikolaou2023fragmentation}.
We adopt the game theoretical approach to decompose all interactions into synchronous bilateral interactions between two agents $i$ and $j$.
The payoff from these interactions depends on the strategies chosen by $i$ and $j$.
$A^{i}(s^{i}_{t},s^{j}_{t})$, the payoff gained by agent $i$ from the interaction with agent $j$, is usually different from $A^{j}(s^{i}_{t},s^{j}_{t})$ gained by agent $j$ from the same interaction.
The possible payoffs are summarized by the payoff matrix of the Prisoner's Dilemma game:
\begin{equation}
	\label{eq:2}
	A^{i}(s^{i}_{t},s^{j}_{t}) =  \begin{bmatrix}
		P & T \\
		S & R
	\end{bmatrix}
\end{equation}
When both $i$ and $j$ defect (i.e., $s^{i}_{t}=0,s^{j}_{t}=0$), both receive $A^{i}(0,0)=A^{j}(0,0)=P$.
If, on the other hand, both cooperate, both receive $A^{i}(1,1)=A^{j}(1,1)=R$.
If, however, only $i$ cooperates while $j$ defects, $i$ receives $A^{i}(1,0)=S$ and $j$ receives $A^{j}(1,0)=T$.

The canonical payoff values $(T,R,P,S)=(5,3,1,0)$ satisfy the inequalities, $T>R>P>S$ and $T-R>P-S$.
They imply that payoffs from defection, $(T,P)$, are always higher than payoffs from cooperation, $(R,S)$.
Since both agents face the same choice, it is rational for both of them to defect.
However, this equilibrium is inefficient because the payoff $P+P$ for both agents is less than $R+R$, the payoff when both agents cooperate.

\paragraph*{Repeated interactions.\ }

In our model, during one time step, each agent $i$ interacts with $n_g$ randomly chosen agents, using the same strategy $s^{i}_{t}$.
Then the accumulated payoff from these interactions is:
\begin{equation}
	\label{eq:10}
	F^{i}_{t} = \sum_{j\in\mathcal{N}_{t}^{i}}A^{i}(s^{i}_{t},s^{j}_{t})
\end{equation}
The set of counterparties denoted as $\mathcal{N}^{i}_{t}$ with $\abs{\mathcal{N}_{t}^{i}}=n_{g}$ changes at every step.
Since these counterparties are randomly sampled from the whole population, the  probability of encountering a cooperator is proportional to the global fraction of cooperators, $f_{t}$.
For $n_{g}$ interactions the expected number of cooperators is thus $n_g f_{t}$
\begin{defn}[Expected payoffs]
The \textit{expected payoffs} for cooperators ($s_t^i=1$) and defectors ($s_t^i=0$) are:
    \begin{align}
	\label{eq:exp-pay-int}
	\mathds{E}[F^{i}_{t}, s^{i}_{t}=1] & =n_g f_{t}R + n_g (1-f_{t})S \nonumber \\
	\mathds{E}[F^{i}_{t}, s^{i}_{t}=0] & =n_g f_{t}T + n_g (1-f_{t})P
\end{align}
where $n_g$ is the number of interactions, $f_t$ is the global fraction of cooperators and $(T,R,P,S)$ are payoff values from the PD game.
\end{defn}
Because of $T>R>P>S$, for any value for $f_{t}$ and $n_{g}$ the expected payoff for \emph{defection} is always greater or equal to the expected payoff for \emph{cooperation}.
If agents choose their strategy $s^{i}_{t}$ such that the expected payoff is maximized,
every agent in the end will defect, i.e. $f_{t}\to 0$ for long times, which is a known, but not desired outcome.

\subsection{The government perspective}
\label{sec:central}

\paragraph*{Identifying cooperators.}

The government has an interest to avoid complete defection.
To increase the fraction of cooperators in the population over time, $f_{t}\to 1$, they are willing to pay a \emph{bonus} $b$ to cooperators.
This, however, requires them to identify the $N^{c}_{t}$ cooperators in a population of $N$ agents.
Here the government relies on some identification technology that correctly identifies agents' strategies only with a probability.
There are indeed technologies to identify defective behavior, e.g. tax evasion or asylum fraud, but we refrain from reducing our general approach to specific technologies. 

\begin{defn}[Identification accuracy]
    We define identification accuracy the probability $p>0.5$ with which the government correctly identifies agents' strategies.
    That means each cooperator correctly receives the bonus with probability $p$ and each defector by mistake receives the bonus with probability $(1-p)$.
    $p<0.5$ would imply that defectors receive a bonus more likely than cooperators, which needs to be avoided.
    $p=0.5$ means that both defectors and cooperators receive a bonus undifferentiated, which renders the policy useless.
    We exclude this case also because derived variables, such as the critical bonus, \cref{eq:bonus_general} contain a term $(2p-1)$ in the denominator and thus would diverge. 
\end{defn}

To model the uncertainty in assigning the bonus $b$, we introduce a binary variable $z^{i}\in\{0,1\}$,  where $z^{i}=1$ indicates that the agent receives the bonus, and 0 otherwise. 
This variable $z^{i}$ follows a Bernoulli distribution $Pr(z^{i}=s)=p^{s} {(1-p)}^{1-s}$. The bonus received at time $t$ by agent $i$ is then given by the random variable $b^{i}(s^{i}_{t}) = bz^{i}$.

\begin{defn}[Expected bonus]
    The \textit{expected bonus} for cooperators and for defectors is:
\begin{align}
	\label{eq:exp-bonus}
	\mathds{E}[b^{i},s^{i}_{t}=1] =bp \quad ; \quad \mathds{E}[b^{i},s^{i}_{t}=0] = b(1-p)
\end{align}
\end{defn}

With the introduction of bonuses, we can define the income that agents expect to receive from their choice.
\begin{defn}[Expected income]
	The \textit{income} $I^{i}_{t}$ of agent $i$ using strategy $s^{i}_{t}$ is defined as the sum of the payoffs from interactions, \cref{eq:exp-pay-int}, and the bonus from the government, \cref{eq:exp-bonus},  in expectation:
	\begin{equation}
		\label{eq:income}
		\mathds{E}[I^{i}_{t},s^{i}_{t}] \equiv I^{i}(s^{i}_{t}) = \mathds{E}[F^{i}_{t},s^{i}_{t}] + \mathds{E}[b^{i},s^{i}_{t}]
              \end{equation}
              This income is subsequently taxed by the government, which reduces the available amount spent by the agents. 
              But in the next time step, agents generate a new income.  
\end{defn}

\paragraph*{Number of bonuses paid.}

Incentivizing cooperators requires the government to secure a budget for paying the bonus.
Given the identification accuracy $p$ for cooperators, only $pN^{c}_{t}$ out of $N^{c}_{t}$ expected cooperators will be correctly identified.
At the same time, there are $N^{d}_{t}=N-N^{c}_{t}$ defectors, and $(1-p)N^{d}_{t}$ defectors will wrongly receive a bonus for cooperation.
\begin{defn}[Expected total number of bonuses]
    The \textit{expected total number of bonuses} paid by the government depends on $p$ as follows:
\begin{align}
    \label{eq:n_bonus}
	\mathds{E}[N^{b},t]\equiv N^{b}_{t} & = N^{c}_{t} p + N^{d}_{t} (1-p) \nonumber \\ &= N [f_{t} p +(1-f_{t})(1-p)]
\end{align}
\end{defn}
$N^{b}_{t}$ equals the expected number of cooperators $N^{c}_{t}=f_{t}N$ only for $f_{t}=0.5$.
If $f_{t}<0.5$, $N^{b}_{t}>N^{c}_{t}$, i.e. the government pays more bonuses than needed because it \emph{overestimates} the number of cooperators if the level of cooperation is \emph{low}.
If $f_{t}>0.5$, on the other hand, $N^{b}_{t}<N^{c}_{t}$, i.e. the government pays less bonuses than needed because it \emph{underestimates} the number of cooperators if the level of cooperation is \emph{high}.

\paragraph*{Income tax.}
To afford paying $N^{b}_{t}$ bonuses, the government needs resources.
These are obtained by means of an \emph{income tax} collected by the government as a fixed fraction $\alpha$ of the income generated by each agent $i$.
Using the simplified notation for expected values, the government can expect a budget from the income tax:
\begin{align}
	\label{eq:6}
	G^{a}_{t}=\alpha \sum_{i=1}^{N}I^{i}_{t}(s^{i}_{t})=\alpha\sum_{i=1}^{N}F^{i}_{t}+ \alpha N^{b}_{t}
\end{align}
Note that at each time step, agents generate a new income, and governments collect new income taxes.
Accumulation of wealth for agents, which could result in additional wealth taxes, is not considered here \citep{Schweitzer-Verginer-Vaccario-2020-should}. 
\paragraph*{Costs.}

Collecting taxes, identifying cooperators, and organizing the payment of bonuses are costly activities for the government.
For the administrative costs we consider a cost function
\begin{align}
	\label{eq:costs}
	C(N^{d}_{t},p,f_{t})=N[c_{r}(p)+(1-f_{t})c_{a}]
\end{align}
that scales with the total number of agents, $N$, and captures two different origins.
The \emph{identification cost} per agent $c_{r}(p)$ is seen as a \emph{direct} cost spent by the government to identify cooperators, e.g. by means  of a technology.
It applies to all agents and increases with the identification effort, or accuracy $p$, for which
different forms are discussed below.
For the government it makes sense to invest into a better accuracy because rewarding defectors by mistake is a wasted effort, or an \emph{inefficiency} inherent to the process.

The second cost arises from the presence of defectors and is therefore
proportional to the level of defection in the agent population, with $c_{a}$ as a parameter.
The government has to \emph{mitigate} the negative impact of defectors, e.g. by paying for additional services that compensate the absence of cooperation.
Hence, we refer to these as \emph{mitigation costs}.

\paragraph*{Budget.}

By combining the income from tax, \cref{eq:income}, the costs, \cref{eq:costs}, and the bonuses paid, \cref{eq:n_bonus}, we define the available budget of the government.
\begin{defn}[Budget of the government]
	The \textit{budget} of the government at time $t$, $G_{t}$, is defined as the sum over all income taxes $G^a_{t}$  minus administrative costs $C(N^{d}_{t},p,f_{t})$ and bonuses paid $N_t^bb$:
	\begin{align}
		\label{eq:budget}
		G_{t} & \equiv  G^{a}_{t}  - C(N^{d}_{t},p,f_{t})  -N^{b}_{t}b \nonumber               \\
		      & = \alpha \sum^{i}F^{i}_{t} - N\left[c_{r}(p)+(1-f_{t})c_{a}\right]  \nonumber \\
		      & \quad - N b (1-\alpha)\left[ f_{t}p + (1-f_{t})(1-p)\right]
	\end{align}
\end{defn}

\begin{figure}
	\centering
	\footnotesize
	\includegraphics[width=0.43\textwidth]{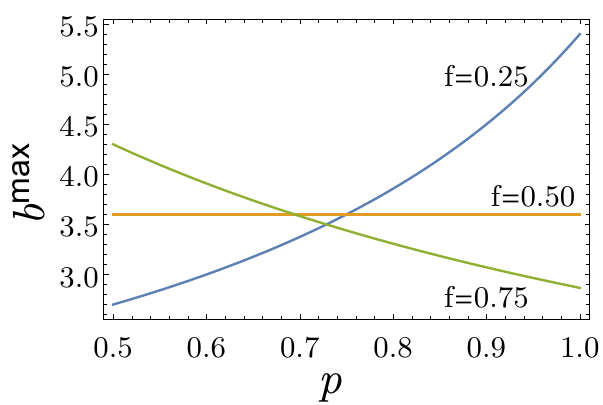}
	\caption{$b^{\mathrm{max}}(f,p)$, \cref{eq:b_max} for $f$=0.25 (blue), 0.5 (orange), and 0.75 (green). Parameters: $n_g=4$, $c_{a}$=0, $c_{r}(p)$=0, $\{T,R,P,S\}=\{5,3,1,0\}$.}
	\label{fig:bmax}
\end{figure}

\subsection{Maximum and minimal bonus.}
\paragraph*{Maximum bonus} When the government spends all of its budget (i.e., $G_{t}=0$), it pays the maximum bonus that can be fully covered by income tax. 
\begin{defn}[Maximum bonus]
\textit{The maximum bonus} that the government can pay is
\begin{align}
	\label{eq:b_max}
	b^{\mathrm{max}}(f_{t},p) = \frac{\alpha \sum^{i} F^{i}_{t} -  N\left[c_{r}(p)+(1-f_{t})c_{a}\right]}{N (1-\alpha)\left[ f_{t}p + (1-f_{t})(1-p)\right]}
\end{align}
\end{defn}

To interpret $b^{\mathrm{max}}(f_{t},p)$, consider the case where the cost function $C(N_{t}^{d},p,f_{t})$ is set to zero.
Then, the maximum  bonus simplifies to
\begin{align}
	b^{\mathrm{max}}(f_{t},p)\big|_{c_{a}=c_{r}(p)=0} = \frac{\alpha \sum^{i} F^{i}_{t}}{N (1-\alpha)\left[ f_{t}p + (1-f_{t})(1-p)\right]}
\end{align}
which still depends on the identification accuracy $p$ because for low values of $p$ many defectors will receive a bonus at the expense of the cooperators.
We would expect that the higher the accuracy $p$ the larger the maximum bonus paid.
Figure~\ref{fig:bmax}, which shows $b^{\mathrm{max}}(f_{t},p)$ as a function of $p$ for different fractions of cooperation $f_{t}$, confirms our expectation only for low levels of cooperation, $f_{t}<0.5$.
Then, indeed,  $b^{\mathrm{max}}(f_{t},p)$ monotonously \emph{increases} with $p$ because the government pays in total fewer bonuses and the bonus per cooperator
increases.
The situation inverts for high levels of cooperation, $f_{t}>0.5$.
Then, $b^{\mathrm{max}}(f_{t},p)$ monotonously \emph{decreases} with  $p$ because there are more cooperators in the population \emph{and} they are correctly identified.
Hence, a smaller bonus per cooperator becomes available.

\paragraph*{Critical conditions for cooperation.} %
Given the setup of the model, the first question regards the minimum bonus needed to incentivize cooperators.
As explained above, without receiving a bonus,
perfectly rational agents would always choose defection.
Hence the minimal bonus has to be high enough to make cooperation as attractive as defection \citep{Schweitzer-Verginer-Vaccario-2020-should}.

\begin{defn}[Difference in expected income]
    We define the difference in expected income for an agent $i$ at time $t$ as
    \begin{equation}
        \mathds{E}[\Delta I_{t}^{i}]\equiv\Delta I_{t}^{i}= I_{t}^{i}(s_{t}^{i}=1) -I_{t}^{i}(s_{t}^{i}=0)
    \end{equation}
\end{defn}
When $\Delta I_{t}^{i}>0$, cooperation is expected to reward agent $i$ with an higher income. When $\Delta I_{t}^{i}<0$ defection is the better choice.

\begin{thm}[Minimal bonus level]\label{min-bon-level}
	Perfectly rational agents are indifferent between choosing cooperation or defection if
	the bonus $b$ reaches the minimum level $b^{*}(f_{t},p)$:
	\begin{equation}
		\label{eq:bonus_general}
		b^{*}(f_{t},p) = \frac{n_g}{2p - 1} [f_{t}(T- R) + (1-f_{t}) (P-S)] \, .
	\end{equation}
      \end{thm}
      The term ``perfectly rational'' refers to an established
      concept in economics and decision theory.
      Agents have a perfect ability to calculate/compare utilities
      and do not make mistakes in maximizing their utility \cite{Eisenführ-Weber-Langer-2010-rationaldecisionmaking}.
Since $T>R>P>S$ and $p>0.5$, $b^{*}(f,p)$ is always positive.
It increases with the number of interactions $n_g$ to compensate for the larger temptation to defect if  more interactions occur.
At the same time, $b^{*}(f,p)$ decreases if the identification accuracy $p$ increases because bonuses are not wasted on wrongly identified cooperators.

\begin{cor}[Critical level of cooperation]\label{cor:crit-f-ratio}
	For a fixed bonus level $b$ and a fixed identification accuracy $p$ a minimum fraction of cooperation $f^*(b,p)$ is needed to equalize the income for cooperators and defectors. This minimum fraction of cooperation is
	\begin{equation}
		\label{eq:f-star}
		f^*(b,p) = \frac{ b (2p - 1) - n_g(P-S) }{ n_g(T- R - P + S)}
	\end{equation}
\end{cor}
Note that the results for the minimal bonus level, \cref{eq:bonus_general}, and for the critical level of cooperation, \cref{eq:f-star}, hold for general payoff values $\{T,R,P,S\}$.
However, our 
calculations for the Figures and for the stability of equilibria, further below, have to assume  fixed payoff values $\{5,3,1,0\}$.

\begin{figure}
	\centering
	\includegraphics[width=0.43\textwidth]{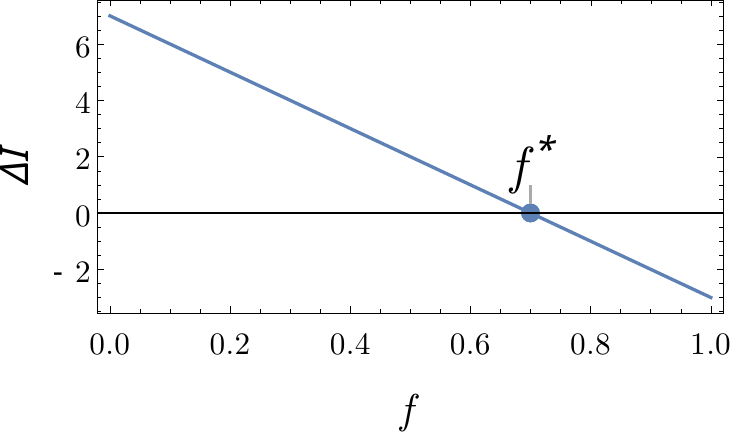}
	\caption{Difference in the expected payoffs $\Delta I$ \cref{eq:cond-2}.
		Parameters: $b$=17, $n_g$=4,, $\{T,R,P,S\}=\{5,3,1,0\}$,  $c_{a}$=0, $c_{r}(p)$=0
		From \cref{eq:f-star}, we find that $f^*$=0.7.}
	\label{fig:deltaI}
\end{figure}

Figure~\ref{fig:deltaI} shows $\Delta I_{t}^{i}$ for changing $f_{t}$, to highlight the minimal level of cooperation, $f^*(b,p)$.

\subsection{Cooperation as evolutionarily stable strategy (ESS)}
\label{sec:cooperation-as-ess}

The above conditions inform about the requirements to incentivize cooperation.
It remains to demonstrate how these conditions impact the evolutionarily stable strategy (ESS) for agents.
At the population level, 
an ESS cannot be displaced by an alternative strategy, i.e. the alternative strategy will not be adopted by a majority of agents (for the Proof see the Appendix).   

\begin{thm}[ESS for perfectly rational agents]\label{th:ESS-rational}
	When bonuses $b$ and accuracy $p$ are fixed, and agents are perfectly rational, their evolutionarily stable strategy is a mixed strategy: they  choose to cooperate with probability $\phi^{*} = f^*(b,p)$ where $f^*(b,p)$ is given by \cref{eq:f-star}.
\end{thm}

\section{Stationary bonus and the cost of irrationality}
\label{sec:study-equilibria}

\paragraph*{Stationary bonus level.}

The above considerations have set two limits for the government to incentivize cooperation.
On the one hand, there is a \emph{minimum} bonus $b^{*}(f_{t},p)$, \cref{eq:bonus_general} that needs to be paid to cooperators to not switch to defection.
This depends on the level of cooperation $f_{t}$: the larger $f_{t}$, the higher the minimum bonus needed to maintain cooperation.
On the other hand, there is a \emph{maximum} bonus $b^{\mathrm{max}}(f_{t},p)$, \cref{eq:b_max}, that the government can only afford to pay, given the income tax collected and the cost structure.

In this Section we explore how these two limit cases relate to the bonus needed to stabilize cooperation in a stationary state with a constant fraction, $f_{t}\to f$.
In a stationary state agents may change their strategies $s^i_t$; however the level of cooperation does not change.
\begin{defn}[Stationary state]
	\label{defn:ss}
	The stationary state is defined by $f_{t+1} = f_{t}\equiv f \,\, \forall t$ where $f$ is the stationary level of cooperation.
\end{defn}
To identify stationary states requires us to define a dynamics for the expected fraction of cooperators.
To define this dynamics, we first define how individual agents change their strategies and then aggregate these changes at the system level.
We have already explored the reference case of \emph{rational} agents that always choose the strategy that maximizes their expected income (see \cref{th:ESS-rational}).
We now focus on agents with \textit{bounded rationality}~\cite{simon1955behavioral} which may choose sub-optimal strategies with a certain probability.
Hence, their decision is modeled with a probabilistic approach.

\subsection{Bounded rationality and transition probabilities}
\label{sec:mean-field-appr-1}

Based on discrete choice theory~\cite{Strzalecki-2025-stochasticchoicetheory},
we model agents' decision to switch their strategy by means of transition probabilities.
In contrasts to perfect
rationality, there is a chance
that agents make mistakes in choosing their best strategy.
This is often denoted as ``bounded rationality'' \cite{Tisdell-2023-advanced}. 
\begin{defn}[Transition probabilities]
We define $w[s|1 - s]$ the transition probability of an agent taking strategy $1-s$ to switch to $s$ as follows \cite{Helbing-1995-quantitativesociodynamics}:  
\begin{align}
	\label{eq:dct}
	w[s|1 - s]
	         & = \frac{
		\exp{\left(\beta  U(s)\right)}}{
		\exp{\left(\beta U(s) \right)} +
	\exp{\left(\beta U(1 - s) \right)  }} \nonumber                     \\
	         & = \frac{1}{1 + \exp{\left(\beta (U(1-s)-U(s))\right) } } 
\end{align}
where $U(s)$ is a utility resulting from choosing strategy $s$, for which different assumptions are discussed below, so no definition is given here.
Agents choose with
a larger probability the strategy $s$ that has a higher
utility, i.e. either $U(s)$ or $U(1-s)$. 
The parameter $\beta$ controls for the rationality of agents~\cite{Bilancini-Boncinelli-Nax-2021-what}.
\end{defn}

For $\beta\rightarrow\infty$, agents will always prefer the strategy yielding the higher utility, no matter how small the difference is.
Hence, this is the limit of perfectly rational agents.
For $\beta\to 0$, on the other hand, agents will choose their strategy more randomly.
Note that all agents know their utilities obtained from both strategies.
Bounded rationality means that they do not always realize this information and instead choose strategies only with a certain probability.
Moreover note that $w[s|1-s]$ are the same for all agents which is possible because agents interact with randomly chosen counterparties.

The transition probabilities allow us to compute the probability that an agent $i$ is cooperating after one time step, i.e., $p[s^{i}_{t+1}=1]$.
To explicitly write this probability, we have to chose whether agents update their strategies synchronously or asynchronously.
The former implies that all agents may change their strategies simultaneously (at every time step), while the latter implies that only one randomly sampled agent may change her strategy (at every time step).
For the synchronous case, 
\begin{align}\label{eq:synch}
p_{\text{Syn}}[s^{i}_{t+1}=1] = p[s^{i}_t=1]w[1|1] + p[s^{i}_t=0]w[1|0]
\end{align}
where the first term is the probability that the agent $i$ was already cooperating and keeps cooperating $w[1|1]=1-w[0|1]$; the second term is the probability that $i$ switches strategy from defection to cooperation.
For the asynchronous case,
\begin{align}\label{eq:asynch}
    p_{\text{Asyn}}[s^{i}_{t+1}=1] = \frac{1}{N}p_{\text{Syn}}[s^{i}_{t+1}=1]+ \frac{N-1}{N}p[s^{i}_t=1]
\end{align}
where the first term is the probability that the agent $i$ is sampled and updated; the second term is the probability that $i$ is not sampled but was already cooperating.

\subsection{Expected dynamics of cooperation.}
\label{sec:mean-field-appr-2}

\begin{thm}[Change of the level of cooperation]\label{th:change-in-f}
The change of the level of cooperation over time is described by the following dynamic
\begin{align}
	f_{t+1}-{f}_{t}=u\Big\{(1-f_t)w[1|0] - f_t w[0|1]\Big\}
	\label{eq:meanfield-f}
\end{align}
where $u=1$ for the synchronous case and $1/N$ for the asynchronous case.
\end{thm}

\begin{cor}[Stationary condition]\label{cor:stat-cond}
Stationary states satisfy the following condition:
\begin{align}
	\label{eq:9}
	\frac{w[1| 0]}{w[0| 1]} = \frac{f}{1-f}
\end{align}
\end{cor}

\subsection{Decisions based on income differences}
\label{sec:decis-based-income}

\begin{thm}[Critical bonus ($\mathcal{I}$)]\label{th:crit-b-I}
	If the utility equals the agent's expected income, $U(s)=I(s)$, \cref{eq:income}, the critical bonus to sustain a stationary fraction of cooperators $f$ is
	\begin{equation}
		\label{eq:quas_stat_I}
		b_{\mathcal{I}}(f,p) =  b^{*}(f,p) + \frac{1}{(2\,p - 1)\beta}\log{\left( \frac{f}{1-f} \right) }
	\end{equation}
	where $b^{*}(f,p)$ is the minimal bonus, \cref{eq:bonus_general}.
\end{thm}
$b_{\mathcal{I}}(f,p)$, \cref{eq:quas_stat_I}, diverges for the limits $f\to 0$, $f\to 1$.
This means that full cooperation or defection are not stationary states.
Also it is interesting to  note that in general $b_{\mathcal{I}}(f,p) \neq b^{*}(f,p)$.
If the government pays a bonus $b_{\mathcal{I}}(f,p)$ larger than $b^{*}(f,p)$,
this implies that $f>(1-f)$, i.e., $f>0.5$.
That means
in the stationary state the majority of agents will always cooperate.
However, if the bonus does not reach the minimal value $b^{*}(f,p)$, the fraction of cooperators becomes small, $f \leq 1/2$.
\begin{cor}[The cost of irrationality]\label{cor:cost-of-irr}
Let \( f > 0.5 \) and \( p \in (0.5,1] \). Then the bonus required to have a stationary level of cooperation $f$ from boundedly rational agents, \( b_{\mathcal{I}}(f, p)\), satisfies
\[
b_{\mathcal{I}}(f, p) > b^{*}(f,p),
\]
where \( b^{*}(f,p) \) is the minimal bonus required for fully rational agents. 
The difference
\[
h(f,p) := b_{\mathcal{I}}(f, p) - b^{*}(f) = \frac{1}{(2\,p - 1)\beta}\log{\left( \frac{f}{1-f} \right) }
\]
represents the additional cost to the government due to bounded rationality.
\end{cor}

The fact that for $f>0.5$, the bonus $b_{\mathcal{I}}(f,p)$, \cref{eq:quas_stat_cond_bonus}, is greater than the minimal  bonus $b^{*}(f,p)$, \cref{eq:bonus_general}, can be interpreted as an additional cost $h(f,p)$ the government has to bear as agents are bounded rational.
For perfectly rational agents (i.e., $\beta\to\infty$), this cost goes to zero, while it diverges for agents choosing strategies more randomly (i.e., $\beta\to 0$).
This cost also diverges for $f \to 1$.
This means  full cooperation in a stationary state is not achievable.
On the other hand, thanks to the bounded rational decisions of agents, it is easier to stabilize a minority of cooperators with $f<0.5$.
In this case, the bonus $b_{\mathcal{I}}(f,p)$ to be paid by the government is less than the minimal bonus $b^{*}(f,p)$.
This implies that any bonus $b<b^{*}(0,p)$ given to rational agents would be wasted as $f\to 0$ in the stationary limit; however, the same bonus $b$ could still maintain a low but non-zero level of cooperation among boundedly rational agents.

\subsection{Decisions based on social welfare differences}
\label{sec:decis-based-welfare}

In scenario $\mathcal{I}$ bounded rational agents are \emph{selfish} because they choose the strategy that maximizes their own income.
We now propose a different scenario, $\mathcal{W}$, in which agents are \emph{social} in that they choose the strategy that maximizes the social welfare of their respective subpopulation.
These subpopulations are defined by the fraction of cooperators and defectors.
\begin{defn}[Social welfare ($\mathcal{W}$)]
	The social welfare of the subpopulation using strategy $s$ is defined as the sum over all expected incomes $I^{i}_{t}(s^{i}_{t})$, \cref{eq:income}, of agents using strategy $s^{i}_{t}=s$.
	\begin{align}
		\label{eq:welfare}
		\mathds{E}[W_{t},s=0]\equiv W(s=0) & =\sum_{i=1}^{N}I^{i}_{t}(s^{i}_{t}=0)\nonumber \\ &= N (1-f_{t}) I(s=0) \nonumber \\
		\mathds{E}[W_{t},s=1]\equiv W(s=1) & =\sum_{i=1}^{N}I^{i}_{t}(s^{i}_{t}=1) \nonumber \\
		                                   & = N f_{t} I(s=1)
	\end{align}
where $I(0)$ and $I(1)$ are the expected incomes from defection and cooperation, respectively.
These expected incomes are the same for all agents as agents interact randomly with any other agents.
\end{defn}
Compared to scenario $\mathcal{I}$, agents still try to choose the strategy that maximizes their utility, but now base their decision on a different information.
Instead of $U(s)=I(s)$, in scenario $\mathcal{W}$ their utility is determined by the expected income of the fraction of cooperators and defectors,  $U(s)\sim W(s)$.
Hence, \emph{differences} in the social welfare of the two subpopulations matter.

\paragraph*{Replicator dynamics.}
For the dynamics we apply the formalism described in \cref{sec:mean-field-appr-1} and \cref{sec:mean-field-appr-2}. 
That means, we can determine the critical bonus needed to sustain cooperation again from \cref{eq:9}, considering that in the transition probabilities, \cref{eq:dct}, the utility now depends on the social welfare, \cref{eq:welfare}.
This will lead us to an expression for the critical bonus $b_{\mathcal{W}}(f,p)$ which has a singularity for $(1-f-p)=0$, i.e. for average $0<f<1$ instead of $f\to 0, 1$ (see \cite{Schweitzer-Verginer-Vaccario-2020-should} for an analysis of this case).
To avoid cases where the critical bonus diverges, we set $U(s)=log(W(s))$ and obtain the following transition rates:
\begin{align}
	w[1|0] & = \frac{(I(1)f_t)^{\beta}}{(I(1)f_t)^{\beta}+(I(0)(1-f_t))^{\beta}}\label{eq:rd-rates-1} \\
	w[0|1] & = \frac{(I(0)(1-f_t))^{\beta}}{(I(1)f_t)^{\beta}+(I(0)(1-f_t))^{\beta}}\label{eq:rd-rates-2}
\end{align}
The role of $\beta$ stays unchanged. 
For $\beta\rightarrow\infty$, agents become rational as they always prefer the strategy yielding the higher utility, no matter how small the difference is.
For $\beta\to 0$, agents choose their strategy more randomly.
Moreover, when using \cref{eq:rd-rates-1} and \cref{eq:rd-rates-2} with $\beta=1$ in \cref{eq:meanfield-f}, we obtain the \emph{replicator equation} for the dynamics of the expected fraction of cooperators~\cite{Schuster-Sigmund-1983-replicator}:
\begin{align}
	f_{t+1}-{f}_{t}=u f_t (1-f_t)  \frac{I(1)-I(0)}{I(1)f_t+I(0)(1-f_t)}
	\label{eq:rd}
\end{align}
where $u=1$ for the synchronous case and $1/N$ for the asynchronous case.
\begin{thm}[Critical Bonus ($\mathcal{W}$)]\label{th:cb-w}
Let \( U(s) = \log(W(s)) \) be the agents’ utility function, where \( W(s) \) denotes the expected welfare of the subpopulation in state \( s \). Then, the critical bonus required to sustain a stationary fraction \( f \) of cooperators, denoted \( b_{\mathcal{W}}(f, p) \), depends on the rationality parameter \( \beta \) as follows:
\begin{equation}
    \label{eq:quas_stat_cond_bonus}
    b_{\mathcal{W}}(f, p)
    \begin{cases}
        < b^{*}(f, p) & \text{if } \beta > 1 \wedge f>0.5, \\
        = b^{*}(f, p) & \text{if } \beta = 1, \\
        > b^{*}(f, p) & \text{if } \beta < 1 \wedge f>0.5,
    \end{cases}
\end{equation}
where \( b^{*}(f, p) \) is the minimal bonus required for fully rational agents, Eq.~\eqref{eq:bonus_general}.
\end{thm}

\begin{cor}[Critical level of cooperation at $\beta=1$($\mathcal{W}$)]\label{cor:crit-f-W}
	Given a critical bonus $b_{\mathcal{W}}(f,p)$, \cref{eq:quas_stat_cond_bonus}, paid by the government to cooperators, and a rationality level $\beta=1$, the expected fraction of cooperators equals $f^*(b,p)$, \cref{eq:f-star}.
\end{cor}

\subsection{Comparing bonus levels}
\label{sec:diff-bonus-levels}

Our approach distinguishes two different types of costs, \cref{eq:costs}: (i) costs to identify cooperators, e.g. by means of a technology, with a parameter $c_{a}$, and (ii) costs to mitigate the impact of defectors, i.e. to compensate the absence of cooperation, with a parameter $c_{r}$.
In the above analysis we have considered these costs in the budget of the government, \cref{eq:budget}.
This has lead us to
the equation for the maximum bonus, $b^{\mathrm{max}}(f_{t},p)$, \cref{eq:b_max}.
It provides a general relation between costs, identification accuracy $p$ and the fraction of cooperators, $f_{t}$.
But $f_{t}$ changes over time, so does the maximum possible bonus.
Therefore, we have first studied the stationary limit, to determine the \emph{critical} bonus needed to stabilize a \emph{stationary} fraction of cooperators.

We have shown that the critical bonus level crucially depends on the information agents take into account for choosing their strategy, therefore we had to distinguish between $b_{\mathcal{I}}(f,p)$, \cref{eq:quas_stat_I}, and $b_{\mathcal{W}}(f,p)$, \cref{eq:quas_stat_cond_bonus}.
In the selfish scenario ($\mathcal{I}$), agents consider only differences in their own income, while in the social scenario ($\mathcal{W}$), they consider differences in the total income, or welfare, of the two subpopulations.
In scenario $\mathcal{I}$ the critical bonus needed to stabilize a given fraction of cooperators was higher than the minimal bonus $b^{*}(f,p)$, \cref{eq:bonus_general} obtained for our reference case of perfectly rational agents. 
The higher bonus arise from cost associated with the irrationality of agents.
In scenario $\mathcal{W}$ the critical bonus may be equal or less than the the minimal bonus $b^{*}(f,p)$ when the rationality of the agents is above a certain threshold.
Thus, we can conclude our findings for the relations between the different bonuses \emph{in the stationary limit} $f_{t}\to f>0.5$ in the following:
\begin{align}
	\label{eq:7}
	b_{\mathcal{W}}(f,p)\leq b^{*}(f,p) \leq  b_{\mathcal{I}}(f,p)\leq b^{\mathrm{max}}(f,p)
\end{align}
where the first inequality holds for $\beta\geq1$ and the last inequality is a necessary condition (rather than a result) to have the government budget to cover for the bonuses.

This brings us to a crucial question: is the government able to afford the critical bonus required to sustain cooperation? 
In the worst-case scenario—where the government exhausts its entire budget—the maximum bonus it can pay is given by $b^{\mathrm{max}}(f,p)$, \cref{eq:b_max}. 
We take this worst case as an upper bound, assuming that the critical bonus exactly matches the government's maximum affordable bonus, leaving no remaining budget. 
Under this assumption, we can analyze the implications in the two different scenarios:
\begin{align}
	\label{eq:stabl_rd}
	b_{\mathcal{W}}^{\mathrm{max}}(f,p) & = b_{\mathcal{W}}(f,p) = b^{*}(f,p)          \\
	\label{eq:stabl_dct}
	b_{\mathcal{I}}^{\mathrm{max}}(f,p) & = b_{\mathcal{I}}(f,p) = b^{*}(f,p) + h(f,p)
\end{align}
where we have assumed $\beta=1$ for the $\mathcal{W}$ scenario as it leads to the replicator equations which also matches the case of having fully rational agents.

\section{Stability of Equilibria}
\label{sec:role-costs}

\subsection{Stability of equilibria (no costs)}
\label{sec:stability-equilibria}

\begin{figure}
	\centering
	\includegraphics[width=0.43\textwidth]{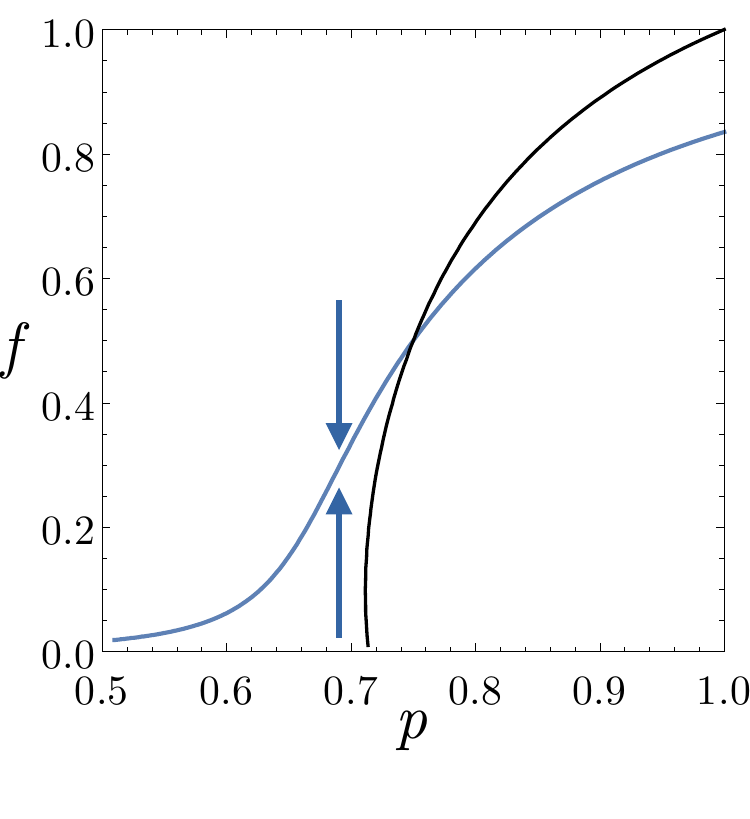}
	\caption{\footnotesize
		$(f,p)$ pairs that solve \cref{eq:stable_f_dct} (blue) and\cref{eq:stable_f_rd} (black).
		The vertical blue arrows indicate the change of $df/dt$ for scenario $\mathcal{I}$ above and below the stationary solution.
		Parameters: $n_g=4$, $\alpha$=$0.4$, $c_{a}$=0, $c_{r}(p)$=0, $\beta=1$,  $\{T,R,P,S\}=\{5,3,1,0\}$.
	}
	\label{fig:bstat_equal_bmax}
\end{figure}

\paragraph*{Endogenous coverage of bonuses. }
From \cref{eq:stabl_rd} and \cref{eq:stabl_dct}, we can derive the $(f,p)$-pairs at which the system reaches a stationary state.
They allow us to study how the identification accuracy $p$ of the government affects the global fraction of cooperators.
We will first investigate the solution in the absence of costs, $c_{a}= c_{r}(p)=0$.
This provides an important reference case to later understand the role of costs, if they were covered from the budget.
The maximum bonus $b^{\mathrm{max}}(f,p)$, \cref{eq:b_max} for scenario $\mathcal{W}$ yields:
\begin{equation}
	\label{eq:stabl}
	\frac{\alpha \sum^{i} F^{i}}{N \left(1 - \alpha \right)[f_{\mathcal{W}}p_{\mathcal{W}}+(1-f_{\mathcal{W}})(1-p_{\mathcal{W}})]} = b^{*}(f_{\mathcal{W}})
\end{equation}
By inserting the canonical payoff values $\{T,R,P,S\}=\{5,3,1,0\}$ in \cref{eq:stabl}, we obtain:
\begin{equation}
	\label{eq:stable_f_rd}
	\frac{n_g\,\alpha\, \left(1 + 3f_{\mathcal{W}} - f_{\mathcal{W}}^2\right)}{(1-\alpha)\left[f_{\mathcal{W}}p_{\mathcal{W}} + (1-f_{\mathcal{W}})(1-p_{\mathcal{W}})\right]} =  \frac{n_g(1+f_{\mathcal{W}})}{2p_{\mathcal{W}}-1}
\end{equation}
The same calculation yields for scenario $\mathcal{I}$:
\begin{multline}
	\label{eq:stable_f_dct}
	\frac{n_g\,\alpha\, \left(1 + 3f_{\mathcal{I}} - f_{\mathcal{I}}^2\right)}{(1-\alpha)\left[f_{\mathcal{I}} p_{\mathcal{I}}  + (1-f_{\mathcal{I}} )(1-p_{\mathcal{I}} )\right]} = \\ \frac{n_g(1+f_{\mathcal{I}} )}{2p_{\mathcal{I}} -1} +\frac{1}{(2\,p_{\mathcal{I}}  - 1)\beta}\log{\left( \frac{f_{\mathcal{I}} }{1-f_{\mathcal{I}}} \right) }
\end{multline}
The $(f,p)$-pairs resulting from \cref{eq:stable_f_dct} are plotted as the blue curve in \cref{fig:bstat_equal_bmax}.
We note that this curve is non-linear, i.e.,
only for $p$ between 0.7 and 0.9, small changes in $p_{\mathcal{I}}$ imply significant changes in $f_{\mathcal{I}}$.

\paragraph*{Stability analysis.}
To prove the stability of the stationary points coming from \cref{eq:stable_f_dct}, we first show that at fixed $p$, there is a unique value of $f$ as solution.
Then, we use the fact that $b^{\mathrm{max}}(f,p)<b_{\mathcal{I}}(f,p)$ above the blue curve in \cref{fig:bstat_equal_bmax} and the change in the fraction of cooperators is expected to be negative.
Below the blue curve, we have instead $b^{\mathrm{max}}(f,p)>b_{\mathcal{I}}(f,p)$ and the change in $f$ is expected to be positive (see the arrows in \cref{fig:bstat_equal_bmax}).
\begin{thm}[Stable equilibria]\label{th:stable-equilibria}
	\label{thm:stable}
	The equilibria satisfying \cref{eq:stable_f_dct} are stable.
\end{thm}
In \cref{fig:bstat_equal_bmax}, we also plot the $(f,p)$ solutions from \cref{eq:stable_f_rd} ($\mathcal{W}$) as a black curve.
One can use a similar approach as above to show that these are stable points.

\subsection{Identifying cooperators in stable states (no costs)}
\label{sec:ident-coop-at}

In addition to the critical bonus paid to sustain a certain level of cooperation, the  government can also obtain information about the accuracy $p$ needed to identify cooperators.
$p$ depends on the identification costs, but also on the information agents use to decide about their cooperation.
Therefore, we continue distinguishing the scenarios $\mathcal{I}$ and $\mathcal{W}$  in the stationary limit.

\begin{cor}[Critical accuracy]\label{cor:crit-p-woc}
	To sustain a stationary level of cooperation $f$ in scenarios $\mathcal{W}$ and $\mathcal{I}$, in the absence of mitigation and identification costs, the government should choose an accuracy
	\begin{align}
		\label{eq:stat-p-rd}
		p_{\mathcal{W}}(f_{\mathcal{W}}) & = \frac{A(f_{\mathcal{W}})}{B(f_{\mathcal{W}})}                                           \\
		p_{\mathcal{I}}(f_{\mathcal{I}}) & = \frac{A(f_{\mathcal{I}}) + C(f_{\mathcal{I}})}{B(f_{\mathcal{I}}) - D(f_{\mathcal{I}})}
	\end{align}
	where
	\begin{align}
		\label{eq:4}
		A(f) & =1 + 3f - f^2 \nonumber                                                        \\
		B(f) & = 1+\alpha-f+7 \alpha f-2 f^2 \nonumber                                        \\
		C(f) & =\frac{1}{\beta n_g}(1-\alpha) (1-f) \log{\left(\frac{f}{1-f}\right)}\nonumber \\
		D(f) & =\frac{1}{\beta n_g}(1-\alpha) (2f-1) \log{\left(\frac{f}{1-f}\right)}
	\end{align}
\end{cor}
We note that $p_{\mathcal{I}}=p_{\mathcal{W}}$ for $f=1/2$.
This property derives from the fact that for $f=1/2$ the cost for irrationality is zero.
\begin{defn}[Accuracy threshold]
	We define the accuracy threshold as
	\begin{align}
		\label{eq:3}
		\hat{p}=p_{\mathcal{I}}(1/2)=p_{\mathcal{W}}(1/2)=1/3 + 1/(6\alpha)
	\end{align}
	Note that $p_{\mathcal{I}}(f_{\mathcal{I}})$ and $p_{\mathcal{W}}(f_{\mathcal{W}})$ are non-decreasing functions in $f$.
	Hence, if $p(f)>\hat{p}$, this implies $f>1/2$.
\end{defn}
\begin{thm}[High cooperation is harder to achieve in scenario $\mathcal{I}$ than in $\mathcal{W}$]\label{th:high-coop-is-hard-in-I}
	In the absense of mitigation and identification costs, if
	the government has choosen an accuracy $p$ greater than the accuracy threshold $\hat{p}$,
	then the stationary level of cooperation in the  $\mathcal{I}$ scenario is lower than in the  $\mathcal{W}$ scenario, i.e., $f_{\mathcal{I}}<f_{W}$.
\end{thm}
As a consequence of the last theorem,
for the remainder of this paper we will focus only on scenario $\mathcal{I}$ because it is more difficult to achieve high cooperation.
Moreover, in scenario $\mathcal{W}$ both the critical bonus and the stationary fraction of cooperators are equal to the case of perfectly rational agents.
In the following, we drop the index $\mathcal{I}$ for simplicity.

\subsection{Impact of mitigation costs}
Now, we consider that from the budget, \cref{eq:budget}, the government must cover also mitigation costs $c_{a}>0$ which increase linearly with the number of defectors, $N (1-f) c_{a}$, \cref{eq:costs}.
\begin{cor}[Increase of identification accuracy ($\mathcal{I}$)]\label{cor:crit-p-wc}
	To sustain a stationary level of cooperation $f$, in presence of mitigation costs but the absence of identification costs the government should choose an accuracy
	\begin{align}
		p(f) = \frac{A(f) + C(f) - \beta c_{a} (1-f) }{B(f) - D(f)- 2 \beta c_{a} (1-f)}\label{eq:stat-p-dct-wac}
	\end{align}
	where $A(f)$, $B(f)$, $C(f)$ and $D(f)$ are given in \cref{eq:4}.
\end{cor}
The mitigation costs have three effects.
First, stationary states (i.e., equilibria) may not exist for low values of $f$.
To see this, note that in \cref{eq:stat-p-dct-wac} the denominator of the stationary accuracy contains negative terms that depend on $c_{a}$ and may make the denominator equal to zero.
For low values of $f$, the accuracy $p$ may exceed the range $(0.5,1]$; hence, stationary states may not exist.
\begin{figure}
	\centering
	\footnotesize
	\includegraphics[width=0.43\textwidth]{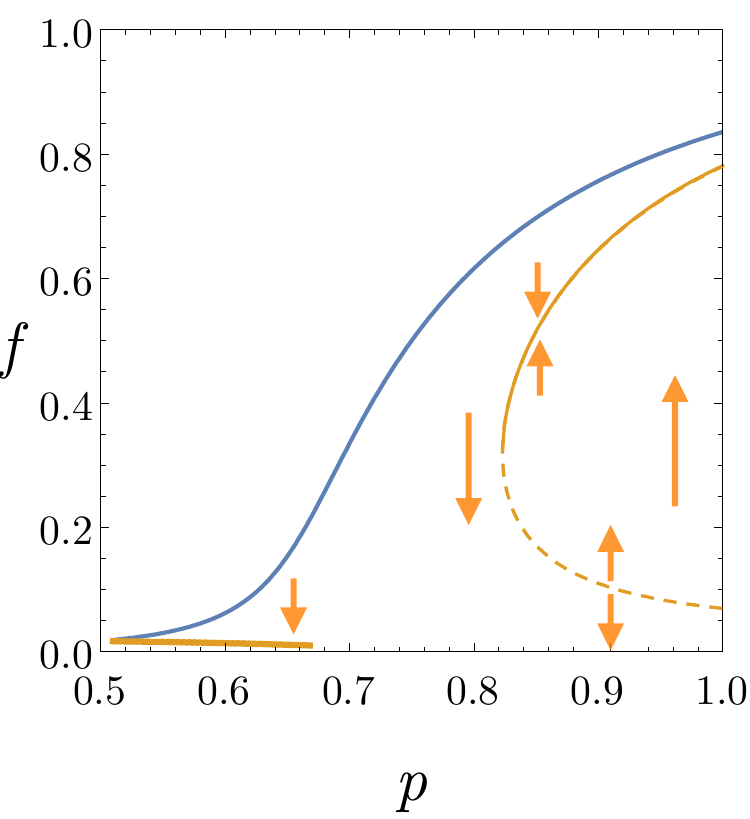}
	\caption{
		$(f,p)$ pairs that solve \cref{eq:stable_f_dct} (blue) and\cref{eq:stat-p-dct-wac} (orange).
		Solid orange lines indicate stable equilibria, dashed ones unstable equilibria.
		The vertical orange arrows indicate the change of $df/dt$ above and below the stationary solution.
		Parameters: $n_g=4$, $\alpha$=$0.4$, $c_{a}$=2, $c_{r}(p)$=0, $\beta=1$,  $\{T,R,P,S\}=\{5,3,1,0\}$.
	}
	\label{fig:bstat_equal_bmax_w_abt_cost}
\end{figure}
To illustrate this, in \cref{fig:bstat_equal_bmax_w_abt_cost},
we have again plotted \cref{eq:stabl_dct} without mitigation costs, $c_{a}=0$, (i.e., the blue curve from \cref{fig:bstat_equal_bmax}) and in orange \cref{eq:stabl_dct} with mitigation costs, $c_a\neq0$.

Second, mitigation costs make it harder to increase the fraction of cooperators.
This result is visible in \cref{fig:bstat_equal_bmax_w_abt_cost} where the area below the blue curve ($c_{a}=0$) is characterized by positive growth, $f_{t+1}-f_t>0$.
In contrast, the area within the orange curve ($c_{a}\neq0$) that has $f_{t+1}-f_t>0$ is much smaller.
Moreover, we find that in the presence of mitigation costs there are  $(f,p)$-pairs at which $f$ \emph{decreases} if $p$ increases (shown by the dashed line in~\cref{fig:bstat_equal_bmax_w_abt_cost}).

Third, the solutions along the dashed line denote unstable equilibria that appear because defectors cause high mitigation costs, and the budget is insufficient to pay the necessary bonuses.
Hence, even for a high identification accuracy $p$, the global fraction of cooperators can  decrease and high cooperation is out of reach.
In conclusion, it is not only crucial to know the stationary states but also under which conditions it is possible to reach them.

\section{The role of identification policies and costs}
\label{sec:accuracy}

In the following we investigate different policies the government may use to adjust the identification accuracy $p$ in response to the fraction of cooperators, $f$.
Then the identification accuracy becomes a variable that the government adjusts over time, $p=p(t)$.
Moreover, we assume a continuous dynamics over time $f_{t+1}-f_t\sim df/dt$.
Note that the conclusions about
the existence of equilibria and their stability, derived below, hold only in this continuous limit.

\subsection{Identification policies (no identification costs)}
\label{sec:exo-ident-costs}

As our reference case, we consider mitigation costs as before, with $c_{a}=3.781$.
But we first analyze policies assuming that \emph{identification} costs $c_{r}$ are \emph{exogenous}, i.e., not paid from the budget.
Therefore, they can be left out from the budget calculations.
But now, instead of focusing on the stationary solutions given by \cref{eq:stat-p-dct-wac} we will consider an explicit dynamics because of the feedback between the fraction of cooperators  $f$ and the adjustment of the accuracy $p$ by the government.
This requires additional assumptions of how $p$ changes with $f$.

\paragraph*{Naive reduction of accuracy. }
First, we assume that $p$ is directly set given the level of cooperation $f$.
One may argue that a high level of cooperation results in a high social welfare, therefore
fewer bonuses are needed to be paid.
A  naive policy may then reduce accuracy when cooperation is high and increase accuracy when cooperation is low.
We formalise this policy with the following definition.
\begin{defn}[Naive policy]
	The government determines the accuracy $p_{t}$ as a function of $f_{t}$,
	\begin{equation}
		\label{eq:naive_p}
		p_{t}\equiv p(f_{t}) = 1 - \frac{f_{t}}{2}
	\end{equation}
	such that $p(0)=1$, while $p(1)=0.5$. %
\end{defn}
To find the stationary level of $f_{t}$ for this policy, we numerically solve the following system of equations for $p(f_{t})$, \cref{eq:naive_p}, $f_{t}$,  \cref{eq:meanfield-f}, $b_{\mathcal{I}}^{\mathrm{max}}(f,p)$, \cref{eq:stabl_dct}.

\begin{figure}[htbp]
	\centering
	\includegraphics[width=0.43\textwidth]{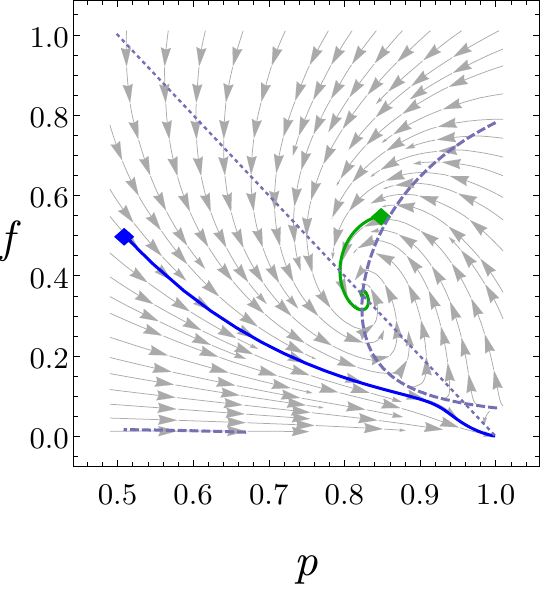}%
	\caption{Naive policy to adjust $p(f)$, \cref{eq:naive_p}: 
          Vector field/stream plot and trajectories $f(p)$ for two two different starting points (diamonds green/blue). Dashed/dotted lines: Changes of $f$ and $p$ are expected to be zero.}
	\label{fig:naive}
\end{figure}
In \cref{fig:naive}, we see that the naive policy leads to two attractors.
The one close to $(f,p)=(0,1)$ means that cooperation disappears because
the budget is used to cover mitigation costs and no sufficient bonuses are paid.
The second attractor has a level of cooperation of about 0.35, reached for an accuracy at about 0.85.
That means the bonus payments are sufficient to reach a stable equilibrium.
We note that the difference in this outcome is entirely determined by the initial conditions.

\paragraph*{Adaptive increase of accuracy. }
As an alternative to the first policy, the government can also adjust the accuracy $p$ depending on the level of cooperation $f$ and a \emph{target cooperation level}, $f_c$.
We formalise this policy as follows.
\begin{defn}[Adaptive policy]
	The government determines the \emph{change} in accuracy $p_{t+1}-p_{t}$ as a function of $f_{t}$,
	\begin{align}
          p_{t+1}& = p_{t} + \mu(f_c - f_{t})
		\label{eq:reactive_p}
	\end{align}
	where $\mu$ is the response parameter. %
\end{defn}
That means the greater the difference between the current and the targeted fraction of cooperators, the larger is the adjustment of the accuracy.
At the same time, when $f_{t}$ is close to $f_c$, the change is small.
\begin{figure}[htbp]
	\centering
	\includegraphics[width=0.43\textwidth]{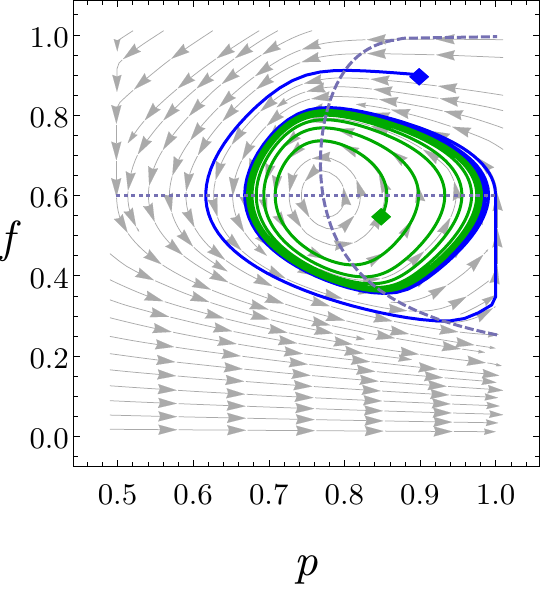}%
	\caption{Adaptive policy to adjust $p(f)$ with $f_c=0.6$, \cref{eq:reactive_p}: 
          Vector field/stream plot and trajectories $f(p)$ for two two different starting points (diamonds green/blue). Dashed/dotted lines: Changes of $f$ and $p$ are expected to be zero.}
	\label{fig:adaptive}
\end{figure}

In \cref{fig:adaptive} we find that the stationary dynamics reaches a limit cycle.
The cycle appears because the stationary state at $f=f_c$ is unstable.
Hence, following this policy may lead to a periodic increase and decrease  in cooperation, while the overall level of cooperation remains between 0.4 and 0.8.

\paragraph*{Enforced increase of accuracy.}
As a
third alternative, the government may instead increase accuracy until full cooperation is reached by setting $f_c=1$, i.e. the government adopts a ``pressing'' policy.
Formally:
\begin{defn}[Pressing policy]
	The government follows the adaptive policy, \cref{eq:reactive_p} with $f_c=1$,
	\begin{align}
		p_{t+1}& = p_{t}+ \mu (1 - f_{t}) \label{eq:pressing_p}
	\end{align}
	such that $p_{t+1}-p_{t}>0$ and $p_{t}$ increases for all $f_{t}<1$.
\end{defn}

\begin{figure}[htbp]
	\centering
	\includegraphics[width=0.43\textwidth]{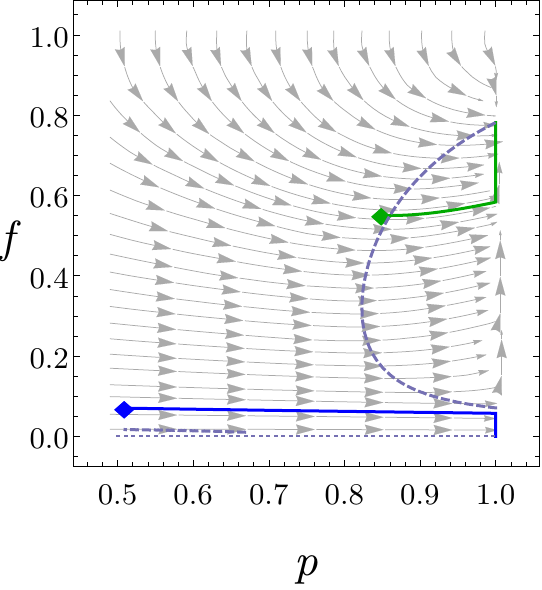}%
	\caption{Pressing policy to adjust $p(f)$ with $f_c=1$, \cref{eq:pressing_p}: 
          Vector field/stream plot and trajectories $f(p)$ for two two different starting points (diamonds green/blue). Dashed/dotted lines: Changes of $f$ and $p$ are expected to be zero}
	\label{fig:pressing}
\end{figure}

In \cref{fig:pressing} we find that a high level of cooperation is obtained, which is also stable.
Moreover, the basin of attraction of this stable point is much larger than for the naive strategy (\cref{fig:naive}).

\subsection{Classification of equilibria}
\label{sec:class-equil}

To formally study the stability of the equilibria points, we compute the Jacobian of the linearized dynamics and its eigenvalues.
According to the standard classification of fixed points  \citep{Strogatz-2000-nonlineardynamicschaos, vulpiani2009chaos},
equilibria become unstable nodes when the eigenvalues are real and positive
and become spirals when the eigenvalues are complex conjugates of each other.

The unstable nodes are found only in a small region of the phase space when $f_c$ is small.
The spirals are stable when the real part of the eigenvalues is negative, which occurs when $f_c>f^{\mathrm{min}}$ where
$f^{\mathrm{min}}$ is the solution of ${dp(f)}/{df}=0$ and $p(f)$ is given by \cref{eq:stat-p-dct-wac}.
The equation ${dp(f)}/{df}=0$ has one solution, which is a minimum, for a sufficiently large values of $c_{a}$.\footnote{If $p(f)$ has no minimum for $f\in[0,1]$, consider $f^{\mathrm{min}}$=0.}

If $f_c<f^{\mathrm{min}}$, spirals are unstable as the real part of the eigenvalues becomes positive.
Limit cycles are attainable in this region. %

\begin{figure}
	\centering
	\includegraphics[width=0.43\textwidth]{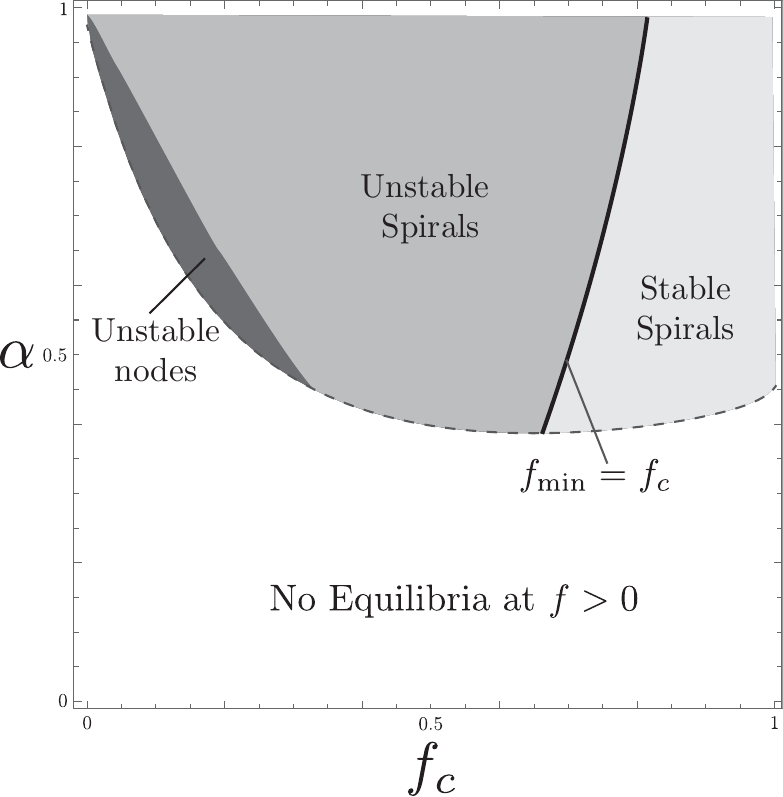}
	\caption{
		Stability of equilibria for the adaptive policies. Parameters:
		$\{T,R,P,S\}=\{5,3,1,0\}$, $\beta=2$ and $c_{a}=3.781$.
	}
	\label{fig:stab-eval}
\end{figure}

In \cref{fig:stab-eval}, we show the stability of the equilibria by utilizing, in addition to the
target fraction of cooperation $f_c$, also the a taxation rate $\alpha$.
According to the budget definition of the government, \cref{eq:budget}, the income tax is the sole source of financing the bonus payments and the identification and mitigation costs.
Hence, the taxation rate needs to reach a minimum level to enable cooperation at all.
This is clearly visible in \cref{fig:stab-eval}.

To obtain a stable level of cooperation, the government must choose a target value $f_c > f^{\mathrm{min}}$.
For example, in \cref{fig:adaptive}(b), \cref{fig:pressing}(b) this is the leftmost point of the dashed curve where $df/dt =0$.
To reach the maximum stable level of cooperation $f$, the government must choose $f_c \geq \max\{\mathcal{F}_p\}$ where $\mathcal{F}_p$ are the values of $f$ solving the equation $p(f)=1$.
If $f_c\in [\min \{\mathcal{F}_p\},f^{\mathrm{min}}]$, the equilibria are unstable.\footnote{An exception is the case when the mitigation costs are too low and $f^{\mathrm{min}}=0$. Then, there is only one value for $f$ such that $p(f)=1$, and there are stable equilibria for $\forall f_c\in[0,1]$.}
Finally, if $f_c < \min \{\mathcal{F}_p\} $, then there are no equilibria for $f\neq 0$.

\subsection{Identification policies (with identification costs)}
\label{sec:endo-ident-costs}

To summarize the previous findings considering mitigation costs $c_{a}$, but without identification costs $c_{r}$, the government can obtain a maximum welfare through a ``pro-active'' policy, i.e., by increasing $p$ when $f$ increases.
This is reasonable because it becomes more attractive to defect if the number of cooperators increases.
Therefore, bonuses must be assigned with higher accuracy to counteract this tendency and ensure that agents keep cooperating.

We wish to compare these baseline findings with a scenario that also considers  \emph{identification} costs.
These are now \emph{endogenous}, i.e. paid from the government's budget, and we assume that higher accuracy is more costly.

\begin{defn}[Identification costs]
	The cost to correctly identify the strategy of one agent with probability $p$ as $c_r(p)$ is defined as
	\begin{align}
		\label{eq:identification_cost}
		c_r(p)= \log{\left(\frac{p}{1-p}\right)}
	\end{align}
\end{defn}
The inverse of this cost function defines an accuracy function:
\begin{align}
	\label{eq:identification_probability_and_cost}
	p(c_r)= \frac{1}{1+e^{-c_r/r}} \,.
\end{align}
where the rescaling factor $r$ is set to one without loss of generality.

This accuracy function satisfies a number of reasonable requirements:
(i) If the government spends nothing to improve accuracy ($c_r=0$), then $p=0.5$, i.e., bonuses are assigned randomly.
(ii) If the government spends more, the accuracy $p$ becomes higher.
I.e., $p(c_r)$ is a monotonously increasing function, ${\partial p(c_r)}/{\partial c_r} > 0 $.
(iii) Increasing accuracy becomes harder if it is already high, ${\partial^2 p(c_r)}/{\partial^2 c_r} < 0 $.
In particular, we have that perfect accuracy becomes infinitely costly, $c_r \to \infty$ if $p(c_r)\to1$.
Note that \cref{eq:identification_cost} is just one of many possible cost functions that  yield the same results discussed below, as long as it satisfies the given requirements.

\begin{figure}
	\centering
	\footnotesize	\includegraphics[width=0.43\textwidth]{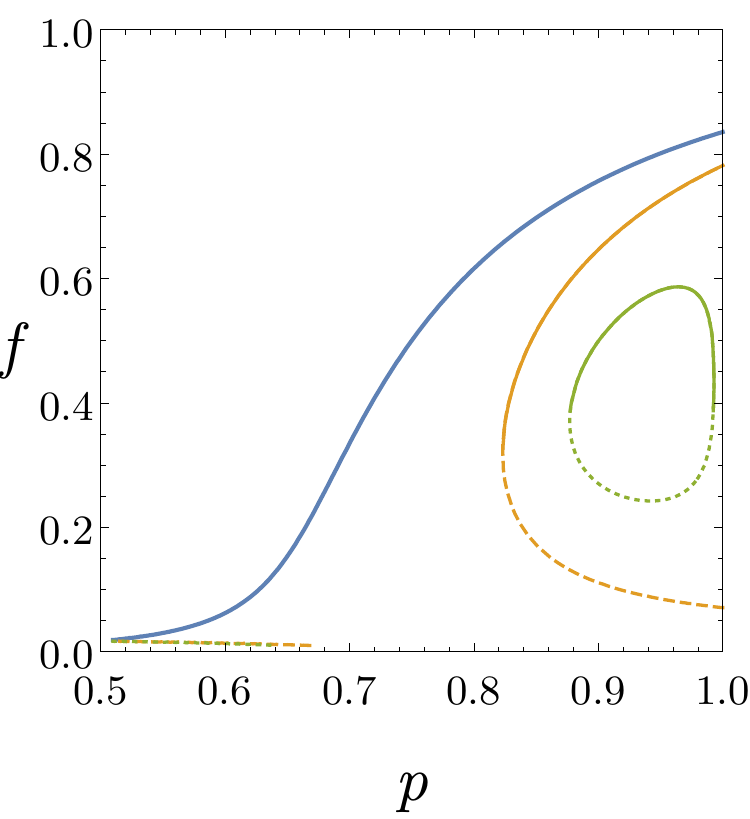}
	\caption{
		$(f,p)$ pairs that solve \cref{eq:stable_f_dct} (blue),\cref{eq:stat-p-dct-wac} (orange) and
		\cref{eq:stabl_dct} (green). Solid lines indicate stable equilibria, dashed ones unstable equilibria.
		Parameters: $n_g=4$, $\alpha$=$0.4$, $c_{a}$=2, $r$=1.6, $\beta=1$,  $\{T,R,P,S\}=\{5,3,1,0\}$.
	}
	\label{fig:p_and_f_w_abt_and_id_cost}
\end{figure}
Identification costs $c_r(p)$ make it harder for the government to increase cooperation.
This leads to new equilibria at $p\lesssim 1$, as the ($f$, $p$)-pairs
in \cref{fig:p_and_f_w_abt_and_id_cost} show.
The blue curve shows the reference case without mitigation and identification costs (see \cref{fig:bstat_equal_bmax}), the orange curve shows the reference case with mitigation costs but without identification costs (see \cref{fig:bstat_equal_bmax_w_abt_cost}), and the green curve considers both costs and spending of the total budget.
I.e. the equilibria are defined by $b^{\mathrm{max}}(f,p)=b_{\mathcal{I}}(f,p)$, \cref{eq:stabl_dct}.
Inside the region defined by the green curve, we have $d f /d t >0$, and
outside this region $df /dt <0$.
Because the green curve is entirely contained within the orange curve, there are $(f,p)$-pairs for which $f$ decreases in the presence of identification costs and increases in their absence.
Moreover, considering identification costs removes the equilibrium points at high values of $p$.
Hence, it is impossible to have (stable) cooperation at $p$=1.

\subsection{Optimal accuracy}
\label{sec:optimal-accuracy}

Even though increasing accuracy makes cooperation more likely, its costs increase and must be covered from the government's budget.
The remaining budget for paying bonuses shrinks, hence there is a tradeoff between increasing accuracy and increasing bonus.
The government needs to make a policy decision based on a utilitarian approach, for instance choosing the ``greatest happiness for the greatest number''.
This means to maximise the  \emph{total payoffs} from interactions, instead of the total income, \cref{eq:income}, or the total welfare, \cref{eq:welfare} which both include bonuses.
In the latter case, maximizing ``happiness''  would be obtained by maximizing bonus payments.
In the former case, however, maximizing total payoffs implies to find the optimal level of cooperation conditional on the costs for accuracy.

\begin{figure}
	\centering
	\footnotesize
	\includegraphics[width=0.43\textwidth]{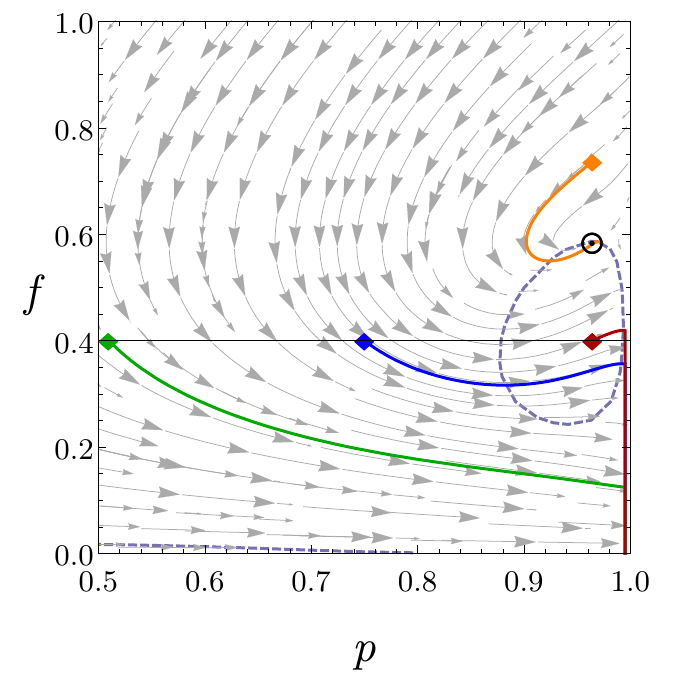}
	\caption{Approach to the optimal final state with the adaptive policy, \cref{eq:reactive_p} and $f_c=f^{\textrm{opt}}$.
	Trajectories for four different starting points are drawn as solid lines.
	The dashed blue line gives the $(f,p)$-pairs satisfying  $b^{\mathrm{max}}(f,p)$ = $b_{\mathcal{I}}(f,p)$. Parameters: $c_{a}$=2, $r$=1.6,  $n_g=4$, $\alpha$=$0.4$, $\beta=1$,$\{T,R,P,S\}=\{5,3,1,0\}$.}
	\label{fig:lagrangian-a}
\end{figure}
\begin{defn}[Optimal cooperation and level for accuracy]
	We define the optimal cooperation fraction $f^{\mathrm{opt}}$ and optimal identification accuracy $p^{\mathrm{opt}}$ as the $(f,p)$-pairs that maximise the total payoff $\sum^{i} F^{i}$, \cref{eq:10}. Formally,
	\begin{equation}
		\label{eq:p_opt_utilitarian_general}
		(f^{\mathrm{opt}}, p^{\mathrm{opt}}) = \argmax_{(f,p)}{\sum^{i} F^{i}}
	\end{equation}
\end{defn}
Among the possible $(f^{\mathrm{opt}},p^{\mathrm{opt}})$-pairs, we are interested in those leading to stationary solutions.
These states satisfy the constraint given by $b_{\mathcal{I}}(f,p)$, \cref{eq:quas_stat_I}.
For the calculation we use the method of Lagrange multipliers to encode the maximisation problem with constraints in one function $\mathcal{L}$.
By setting the gradient of $\mathcal{L}$ equal to zero, we solve the maximisation problem.
As before, we use the assumption that the total budget is spent, hence the bonus paid is the maximum bonus available $b^{\mathrm{max}}(f,p)$, \cref{eq:b_max}.
Then we have:
\begin{equation}
	\label{eq:lagrangian_general}
	\mathcal{L}(f, p, \lambda) =\sum^{i} F^{i} - \lambda \Big[b^{\mathrm{max}}(f,p)-b_{\mathcal{I}}(f,p)\Big] \end{equation}
By inserting the canonical payoffs, we obtain:
\begin{multline}
	\label{eq:lagrangian}
	\mathcal{L}(f, p, \lambda) =N\,n_g \left(1 + 3f - f^2\right)- \\
	\lambda \left(\frac{  \alpha n_g\,\left(1 + 3f - f^2\right) - c_{a}(1-f)- c_r(p)}{(1-\alpha)\left[fp + (1-f)(1-p)\right]} \right. \\ \left. -\frac{n_g\beta(1+f) + \log{\left( \frac{f}{1-f}\right)}}{ (2\,p - 1)\beta} \, \right)
\end{multline}

\begin{figure}
	\centering
	\footnotesize
	\includegraphics[width=0.43\textwidth]{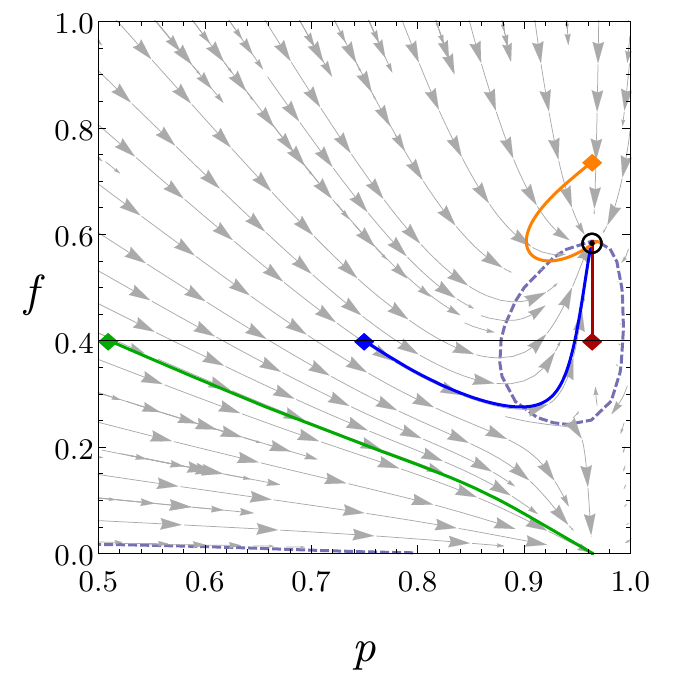}
	\caption{Approach to the optimal final state with the optimal accuracy  policy, \cref{eq:11}.
		Trajectories for four different starting points are drawn as solid lines.
		The dashed blue line gives the $(f,p)$-pairs satisfying  $b^{\mathrm{max}}(f,p)$ = $b_{\mathcal{I}}(f,p)$. Parameters: $c_{a}$=2, $r$=1.6,  $n_g=4$, $\alpha$=$0.4$, $\beta=1$,$\{T,R,P,S\}=\{5,3,1,0\}$.
	}
	\label{fig:lagrangian-b}
\end{figure}

We visualize the problem of the Lagrangian multipliers and the solutions in \cref{fig:lagrangian-a} and \cref{fig:lagrangian-b}.
The stationarity constraint is satisfied along the dashed curve.
The objective function in \cref{eq:lagrangian} is monotonously increasing for $f\in[0,1]$.
Hence, the solution to the Lagrangian problem is the point of maximum $f$ along the dashed curve (the black circle in \cref{fig:lagrangian-a} and \cref{fig:lagrangian-b}).

To reach the optimal level of cooperation,
one could argue that government should choose the adaptive policy, \cref{eq:reactive_p}, with $f_c=f^{\mathrm{opt}}$.
Unfortunately, this is wrong.
In \cref{fig:lagrangian-a} we show that most trajectories starting close to $(f^{\mathrm{opt}}, p^{\mathrm{opt}})$ will end in $f=0$ instead of the optimal point.
For a given initial $f=0.4$, we have chosen three different initial values of $p$, and the blue, green and red trajectories all approach $f=0$.
Only the orange trajectory, which starts from the same $p$, but a higher $f$, converges to the optimal solution.

The convergence to $f=0$ happens because the adaptive policy has a hyperbolic fixed point (i.e., a saddle) in $(f^{\mathrm{opt}}, p^{\mathrm{opt}})$.
This instability arises from the fact that increasing $p$ beyond $p^\mathrm{opt}$ does not increase $f$.
On the contrary, too much is spent on identification accuracy and not enough budget is left to pay bonuses.

To solve this problem, the government should avoid an accuracy  $p>p^{\mathrm{opt}}$.
Therefore we propose the following identification policy.
\begin{defn}[Optimal accuracy policy]
	If the government knows $p^{\mathrm{opt}}$, it can adjust its identification accuracy $p$ such that \begin{equation}
		\label{eq:11}
		\frac{dp}{dt} = \mu (p^{\mathrm{opt}}-p)
	\end{equation}
\end{defn}

This policy differs from \cref{eq:reactive_p}, where the government changes $p$ to reach a target level of cooperation, $f_c$.
When adopting the new policy, \cref{eq:11}, the government may reach $p^{\mathrm{opt}}$ if the initial level of cooperation, $f$, is sufficiently high (see blue line in \cref{fig:lagrangian-b}).
Note that the initial choice of $p$ affects whether and how long it takes to reach the optimal state.
At fixed $f$, the $p$ which minimizes the time to reach the optimal state is $p^{\mathrm{opt}}$ (see red line in \cref{fig:lagrangian-b}).
If the government, however, had set the initial $p$ too low, then it may be impossible to reach the stable points with $f>0$ (see green line in \cref{fig:lagrangian-b}).

As an important insight we find that the total payoff and the  fraction of cooperators are both maximized at accuracy values $p<1$.
That means, the level of cooperation cannot be further increased because the costs of correctly identifying one more cooperator is higher than the benefit.

\section{Discussion and conclusions}
\label{sec:discussion}

\subsection{Model contributions}
\label{sec:model-contributions}

We first summarize the methodological contributions of our paper.
Following \citep{Schweitzer-Verginer-Vaccario-2020-should}, we combine an agent perspective, i.e. the  game theoretical setup of a  repeated prisoner's dilemma, with a governmental perspective that reflects budgeting constraints for interventions.
The underlying assumption is that the government wishes to increase the fraction of cooperators in the population and is willing to pay bonuses for cooperating behavior.
Several costs are thereby incurred : (i) Costs to pay the bonuses, (ii) costs for identifying who shall receive the bonus, and (iii) costs for mitigating the consequences of non-cooperative behavior.
These costs are covered by a governmental budget allocated from taxes on the payoffs agents receive from their repeated interactions.
This way, our model establishes a feedback between agents' individual strategic choices and a governmental policy design at the systemic level.

The novel element is the consideration of different costs and their resulting constraints on the feasible governmental policies.
First, we make explicit that defection has a negative societal impact that need to be mitigated by the government.
That means, the government has to compensate the lack of cooperation, e.g., by paying for extra services.
Hence, cooperation can \emph{save} public spending, while defection causes costs.
This feedback underpins the motivation of the government to increase cooperation.
Second, we make explicit that governmental intervention policies are fallible, by considering an identification accuracy $p$ and identification costs $c_{r}(p)$. %
Modeling the fallibility of incentive policies was not proposed before.
As we demonstrate, naive approaches to increase the identification accuracy may fail to also increase the level of cooperation.
I.e., they only incur costs whereas optimal policies can maximize cooperation without maximising identification accuracy.

\subsection{Policy insights}
\label{sec:policy-insights}

\textbf{Not acting is not a choice.}
Governmental projects to steer public or private decisions toward a preferred outcome have become quite popular, in particular with respect to climate or energy issues.
Examples are subsidies like the ``Superbonus'' in Italy~\citep{eu_superbonus_project} or ``HOMES'' program in USA~\citep{inflation_reduction_act_2022}.
This requires governments to understand, and to manage, the complex interplay between individual interests and societal goals.
Theoretical reflections on this issue also need to address the implicit costs \cite{Hunt-Saunders-Kilmer-2018-estimateslaw} of such subsidies resulting from  administration, monitoring, and fraud detection.
Even more, they have to count in the governmental costs for not improving now, paying instead to compensate for the damage caused by defective behavior.
The negative externalities caused by defection effectively reduce the budget of the government.
With our explicit consideration of identification and mitigation costs, we provide a formal approach to address this problem.

\textbf{Governmental responsibilities.}
While
interventions are costly, it is important to make them \emph{endogenous} rather than exogenous,  i.e. to cover these costs from  the governmental budget in a self-sustaining manner.
In our model, we have captured this requirement in a tax scheme that covers the bonus payments for cooperators.
This taxation creates a tight coupling between total welfare and payable incentives.
If cooperation is too low, it becomes impossible for the government to cover costs \emph{and} bonuses.
Hence, there is a tipping point, at which the system collapses and no cooperation is possible.
From a policy perspective, reaching a critical level of cooperation should be prioritized.

The government is also responsible for a fair distribution of these bonuses, i.e. it needs to provide identification policies which are costly.
Hence there is a tradeoff between accuracy and efficiency to reach the goal of improving cooperation.
In our model, we address different policies to balance between costs and targeted cooperation.
It takes explicitly into account that governmental policies are fallible, a problem often overlooked.

\textbf{More $\neq$ better.}
Our findings refute simplifying beliefs that more effort in identifying cooperative and defective behavior may lead to higher levels of cooperation.
One of the reasons is the greater temptation to defect in cooperative environments, which eventually lead to an increase in the critical bonus payments.
We show that, given the cost structure assumed for the government, perfect cooperation becomes impossible because of infinite costs.
Also, certain identification policies lead to unstable or oscillating levels of cooperation.
We demonstrate how such policies can be optimized if  ``achievable'' and not  ``desirable'' is the ultimate goal for the government.

\textbf{Social welfare instead of income.}
In our model, we demonstrate that the information agents take into account for their strategic choice, has a considerable impact on the critical bonus level and, hence, on the feasibility of governmental incentive policies.
It becomes easier to establish a target level of cooperation if agents decide based on the social welfare of the respective subpopulations, rather than on their individual income.
Thus, promoting this information publicly, is an additional measure for the government to improve cooperation.
Our model also shows that bounded rationality in agents' decisions results in an extra ``cost of irrationality'', which increases the critical bonus level needed.
In other words, sustaining high cooperation is easier among rational agents.
While in the plain prisoner's dilemma defection is the rational choice, we show that, in the presence of incentives, rationality  drives cooperation and system change.

\appendix
\section{List of proofs}

\begin{proof}[Proof of \cref{min-bon-level}]
	Perfectly rational agents try to maximize their income $I_{t}^{i}(s^{i}_{t})$, \cref{eq:income}.
	They choose strategy $s^{i}_{t}$ over $1 - s^{i}_{t}$ when $I_{t}^{i}(s^{i}_{t})$ is expected to be larger than $I_{t}^{i}(1-s^{i}_{t})$, i.e., the expected difference in income $I_{t}^{i}(s_{t}^{i}) -I_{t}^{i}(1-s_{t}^{i})>0$.
    When the difference in expected income is zero, i.e., $\Delta I_{t}^{i}=0$, agent $i$ is indifferent between cooperation and defection.
    In other words, the condition for indifference is:
	\begin{multline}
		\label{eq:cond-1}
		\mathds{E}[F_{t}^{i},s_{t}^{i}= 1] + \mathds{E}[b^{i},s^{i}_{t}=1] = \\ \mathds{E}[F^{i}_{t},s^{i}_{t}=0] + \mathds{E}[b^{i},s^{i}_{t}=0]
	\end{multline}
	For fixed values of $f_{t}$ and $b$, \cref{eq:cond-1} can be rewritten using \cref{eq:exp-pay-int} and \cref{eq:exp-bonus}:
	\begin{align}
		\label{eq:cond-2}
		n_g f_{t}R + n_g (1-f_{t})S + bp = \nonumber \\  n_g f_{t}T + n_g (1-f_{t})P + b(1-p)
	\end{align}
	Solving \cref{eq:cond-2} for $b$, we obtain the minimal bonus $b^{*}(f,p)$,  \cref{eq:bonus_general}, needed to equalize the expected income from defection and cooperation.
\end{proof}

\begin{proof}[Proof of \corref{cor:crit-f-ratio}]
It is sufficient to solve \cref{eq:bonus_general} for $f(b,p)$ and to recall that $\Delta I_{t}^{i}=0$ when $b=b^{*}(f_{t},p)$.
\end{proof}

\begin{proof}[Proof of \cref{th:ESS-rational}]
    To simplify the notation, let us write $f^*(b,p)\equiv f^{*}$. 
	Consider $N-1$ agents cooperating  with probability $\phi^{*}$, while one agent cooperates with probability $\phi^{*}+\epsilon$ with $\epsilon\in[-\phi^{*}, 1-\phi^{*}]$. {This interval for $\epsilon$ ensures that the probability of cooperating, $\phi^{*}$, stays bounded between 0 and 1.}
	The expected number of cooperators is then $(N-1)\phi^{*} + \phi^*+\epsilon$ and, hence, the expected fraction of cooperators is $f=\phi^{*} +\epsilon'=f^*+\epsilon'$, with the rescaled variable $\epsilon'=\epsilon/N$.

	For the $(N-1)$ agents cooperating with probability $\phi^*$, the expected income is
	\begin{align}
		\label{eq:income-others}
		\mathds{E}[I_{\phi^*}] & =\phi^*\,I[f^*+\epsilon',1] + (1-\phi^*)\,I[f^*+\epsilon',0] \nonumber \\  &= I[f^*+\epsilon',0] + \phi^* \,\Delta I[f^*+\epsilon']
	\end{align}
	where
	$I[f, s]$ is the expected income for strategy $s$, given the expected fraction of cooperators, $f$, and $\Delta I[f] = I[f,1] - I[f,0]$.
	For the one agent cooperating with probability $\phi^*+\epsilon$, the expected income is instead:
	\begin{align}
		\label{eq:income-self}
		\mathds{E}[I_{{\phi^* + \epsilon}}] & = (\phi^*+\epsilon)\, I[f^*+\epsilon',1] + \nonumber                 \\
		                                    & \quad + (1-\phi^*-\epsilon)\, I[f^*+\epsilon',0] \nonumber           \\
		                                    & = I[f^*+\epsilon',0] + (\phi^*+\epsilon)\,\Delta I(f^*+\epsilon')\,.
	\end{align}
	Cooperating with probability $\phi^*$ is an evolutionarily stable strategy if $\mathds{E}[I_{\phi^*}] > \mathds{E}[I_{\phi^* +\epsilon}] \,\, \forall\epsilon\in[-\phi^*, 1-\phi^*]$.
	Substituting \cref{eq:income-others} and \cref{eq:income-self} into this inequality leads to:
	\begin{align}
		\label{eq:ess}
		\phi^*\, \Delta I[f^*+\epsilon']
		 & > (\phi^*+\epsilon)\, \Delta I[f^*+\epsilon'] \nonumber \\  \epsilon\,\Delta I[f^*+\epsilon'] & <0
	\end{align}
	To verify \cref{eq:ess}, we show that $\epsilon$ and $\Delta I[f^*+{\epsilon'}]$ always have  opposite signs.
	If $\epsilon$ is positive, then $\Delta I[f^*+{\epsilon'}]$ is negative by definition of $f^*$ (see \cref{eq:f-star} and \cref{fig:deltaI}).
	On the other hand, when $\epsilon$ is negative, $\Delta I[f^*+{\epsilon'}]$ is positive.
	Hence, when all other agents play the mixed strategy $\phi^*$, the agent's best response is to cooperate with probability $\phi^*$.
	It follows that cooperating with probability $\phi^*$ is an evolutionarily stable strategy.
\end{proof}

\begin{proof}[Proof of \cref{th:change-in-f}]
    We start providing proof for the synchronous case.
    From the definition of the expected fraction of cooperators, \cref{eq:5}, computed at time $t+1$, we have $f_{t+1} = \frac{1}{N}\sum_{i}^N p[s_{t+1}^i=1]$. Substituting in this equation \cref{eq:synch}, we obtain:
    \begin{align}
        f_{t+1} &= \frac{1}{N}\sum_{i}^N p[s^{i}_t=1]w[1|1] + p[s^{i}_t=0]w[1|0] \\ \nonumber
                &= f_t w[1|1] + (1-f_t)w[1|0]\\ \nonumber
                &= f_t - f_t w[0|1] + (1-f_t)w[1|0] 
    \end{align}
where the last equality is obtained by recalling that $w[1|1]=1-w[0|1]$. 
By reordering the terms, one obtains \cref{eq:meanfield-f} with $u=1$. 
The proof for the asynchronous case follows the same steps.
\end{proof}

\begin{proof}[Proof of \corref{cor:stat-cond}]
    Recalling the definition of stationary states, i.e., $f_{t+1}=f_t=f$, it is sufficient to impose the right hand side of \cref{eq:meanfield-f} equal to zero.
\end{proof}

\begin{proof}[Proof of \cref{th:crit-b-I}]
	This result follows directly from the stationary state condition, \cref{eq:9}, using \cref{eq:dct} for the transition probabilities and  $U(s)=I(s)$, \cref{eq:income} for the utilities.
\end{proof}
\begin{proof}[Proof of \corref{cor:cost-of-irr}]
	It is sufficient to consider $b_{\mathcal{I}}(f,p)$, \cref{eq:quas_stat_cond_bonus}.
	Because $b_{\mathcal{I}}(f,p)\sim b^{*}(f,p) + \log{\left(  f /(1-f)\right)}$, for $f>0.5$ we have $\log{\left(  f /(1-f)\right)}>0$ and $b_{\mathcal{I}}(f,p) > b^{*}(f,p)$.
\end{proof}
\begin{proof}[Proof of \cref{th:cb-w}]
    Using the stationary condition, \cref{eq:9} and inserting the transition rates for the $\mathcal{W}$ scenario, \cref{eq:rd-rates-1} and \cref{eq:rd-rates-2}, we obtain:
    \begin{equation}
        \left(\frac{W(1)f}{W(0)(1-f)}\right)^\beta = \frac{f}{1-f}\Rightarrow \left(\frac{I(1)}{I(0)}\right)^\beta =  \left(\frac{1-f}{f}\right)^{\beta-1}
    \end{equation}
    For $\beta=1$, we get $I(1)/I(0)=1$. 
    Recalling that $b^{*}(f, p)$ equalizes incomes, \cref{eq:bonus_general}, it implies $b_{\mathcal{W}}(f, p)=b^{*}(f, p)$.
    For $\beta>1  \wedge f>0.5 $,  $((1-f)/f)^{\beta-1}<1$ that implies $I(1)/I(0)<1$. 
    Recalling that $b^{*}(f, p)$ is the minimal bonus to make expected income from cooperation equal the one from defection, we get $b_{\mathcal{W}}(f, p)<b^{*}(f, p)$.
    For $\beta<1  \wedge f>0.5 $,  $((1-f)/f)^{\beta-1}>1$ that implies $I(1)/I(0)>1$ and hence, $b_{\mathcal{W}}(f, p)>b^{*}(f, p)$.
\end{proof}
\begin{proof}[Proof of \corref{cor:crit-f-W}]
	This follows directly from \cref{th:cb-w} and the expression for $b^{*}(f,p)$, \cref{eq:bonus_general}.
\end{proof}

\begin{proof}[Proof of \cref{th:stable-equilibria}]
    To simplify the notation, we set $f\equiv f_\mathcal{I}$.
    We first show that existence and uniqueness of the equilibria, i.e., one solution $f\in(0,1)$ for \cref{eq:stable_f_dct} exists at fixed $p\in(0.5,1]$.
    Then, we show that this solution is stable.
    \textbf{Existence and uniqueness.} Let $p\in(0.5,1]$. 
    Then the left hand side (LHS) of \cref{eq:stable_f_dct} is a (i) continuous (ii) monotonous strictly \textit{decreasing} function taking (iii) positive finite values for $f\in(0,1)$.
    To show (i) and (iii) is sufficient to note that both the numerator and denominator of LHS are continuous function with positive values for $f\in(0,1)$.
    To show (ii), one can take the derivative of the LHS and see that it is always negative for $f\in(0,1)$.
    The right hand side (RHS) instead is a continuous monotonous strictly \textit{increasing} function from $-\infty$ to $+\infty$ for $f\in(0,1)$.
    To show these properties is sufficient to note that all the terms in RHS are continuous monotonous strictly {increasing} for $f\in(0,1)$.
    These two observation implies that we can define a function $g = \mathrm{LHS} - \mathrm{RHS}$ that is continuous monotonous strictly decreasing taking values from $+\infty$ to $-\infty$ for $f\in(0,1)$.
    For the Bolzano's theorem, $g$ has a root in this interval.
    Since $g$ is monotonous strictly decreasing, the root is unique. We name the root $\hat{f}$ and note that $\sgn(g)=\sgn(\hat{f}-f) \forall f\in(0,1)$.
	\textbf{Stability.} $\hat{f}$ is a stable solution if the change $f_{t+1}-f_t$, \cref{eq:meanfield-f}, is positive in the interval $(0, \hat{f})$  (below the blue curve in \cref{fig:bstat_equal_bmax}) and negative in $(\hat{f},1)$ (above the blue curve in \cref{fig:bstat_equal_bmax}).
	To find the sign of $f_{t+1}-f_t$ in these two intervals,
    we note that $\sgn{(f_{t+1}-f_t)} = \sgn(g)$. 
    This follows from the fact that when government pays a bonus $b^{\mathrm{max}}>b_\mathcal{I}$ (i.e., $g>0$), the expected change in cooperation level is positive. 
    When the paid bonus $b^{\mathrm{max}}<b_\mathcal{I}$ (i.e., $g<0$), the expected change in cooperation level is negative.
    \textit{Intuitively}, this is the case as the bonus paid is either higher or lower compared to the critical bonus needed to keep the cooperation fraction constant.
    Hence, the cooperation fraction either increases or decreases.
    \textit{Formally}, it is sufficient to look at \cref{eq:meanfield-f} and verify that $\sgn{(f_{t+1}-f_t)}>0 \, \forall b>b_\mathcal{I}$ and $\sgn{(f_{t+1}-f_t)}<0 \, \forall b<b_\mathcal{I}$.
    Given $\sgn(\hat{f}-f)$=$\sgn{(f_{t+1}-f_t)}$=$ \sgn(g)\forall f\in(0,1)$, we have that $f$ increases over time for $f<\hat{f}$ while decreases when $f>\hat{f}$.
	This implies that $\hat{f}$ is stable.
\end{proof}

\begin{proof}[Proof of \corref{cor:crit-p-woc}]
	It is sufficient to solve \cref{eq:stable_f_rd} and \cref{eq:stable_f_dct} with respect to $p$ and to recall that these equations are satisfied in a stationary stable state.
\end{proof}

\begin{proof}[Proof of \cref{th:high-coop-is-hard-in-I}]
	We show that $f_{\mathcal{W}} = f_{\mathcal{I}} + \epsilon$ with $\epsilon >0$.
	If $p_{\mathcal{I}}>\hat{p}$ and $p_{\mathcal{W}}>\hat{p}$, then  $f_{\mathcal{I}}>1/2$ and  $f_{\mathcal{W}}>1/2$.
	When $f_{\mathcal{I}}\in(1/2,1)$, we have that $C(f)>0$ and $D(f)>0$, \cref{eq:4}, and hence, $p_{\mathcal{I}}(f_{\mathcal{I}})>p_{\mathcal{W}}(f_{\mathcal{W}})$.
	In other words, $\forall f>1/2 \quad \exists \epsilon >0 : p_{\mathcal{I}}(f_{\mathcal{I}}) = p_{\mathcal{W}}(f_{I}+\epsilon)$.
	Hence, $f_{\mathcal{W}} = f_{\mathcal{I}}+\epsilon$ with $\epsilon>0$.
\end{proof}
\begin{proof}[Proof of \corref{cor:crit-p-wc}]
	It is sufficient to solve \cref{eq:stabl_dct} with respect to $p$ and to recall that this equation is satisfied in a stationary state.
\end{proof}

	{
          \setlength{\bibsep}{1pt}
          \small 
          \bibliography{arxiv}
	}

\end{document}